\documentclass[fleqn,usenatbib]{mnras}

\usepackage{newtxtext,newtxmath}

\usepackage[T1]{fontenc}
\usepackage{ae,aecompl}

\usepackage{tikz}
\usetikzlibrary{shapes.geometric, arrows}
\usepackage{lipsum}

\newcommand{\D}{\partial}
\newcommand{\DD}{\frac}

\newcommand{\beq}{\begin{equation}}
\newcommand{\eeq}{\end{equation}}
\newcommand{\ben}{\begin{enumerate}}
\newcommand{\een}{\end{enumerate}}
\newcommand{\bit}{\begin{itemize}}
\newcommand{\eit}{\end{itemize}}
\newcommand{\barr}{\begin{array}}
\newcommand{\earr}{\end{array}}
\newcommand{\bc}{\begin{center}}
\newcommand{\ec}{\end{center}}
\usepackage{graphicx}	
\usepackage{amsmath}	
\usepackage{amssymb}	

\title[Time-implicit schemes in fluid dynamics?]{Time-implicit schemes in fluid dynamics? -- Their advantage in the regime of ultra-relativistic shock fronts}

\author[M. S. Fischer et al.]{
Moritz S. Fischer$^{1}$\thanks{E-mail:  moritz.fischer@uni-hamburg.de},
A. A. Hujeirat$^2$
\\
$^{1}$Hamburger Sternwarte, Gojenbergsweg 112, D-21029 Hamburg, Germany\\
$^{2}$IWR - Interdisciplinary Center for Scientific Computing, Heidelberg University, INF 368, D-69120 Heidelberg, Germany\\
}

\date{Accepted XXX. Received YYY; in original form ZZZ}

\pubyear{2020}

\begin{document}
\label{firstpage}
\pagerange{\pageref{firstpage}--\pageref{lastpage}}
\maketitle

\begin{abstract}
Relativistic jets are intrinsic phenomena of active galactic nuclei (AGN) and quasars. They have been observed to also emanate from systems containing compact objects, such as white dwarfs, neutron stars and black hole candidates. The corresponding Lorentz factors, $\Gamma$, were found to correlate with the compactness of the central objects. In the case of quasars and AGNs, plasmas with $\Gamma$-factors larger than $8$ were detected. However, numerically consistent modelling of propagating shock-fronts with $\Gamma \geq 4$ is a difficult issue, as the non-linearities underlying the transport operators increase dramatically with $\Gamma$, thereby giving rise to a numerical stagnation of the time-advancement procedure or alternatively they may diverge completely.
In this paper, we present a unified numerical solver for modelling the propagation of one-dimensional shock fronts with high Lorentz factors. The numerical scheme is based on the finite-volume formulation with adaptive mesh refinement (AMR) and domain decomposition for parallel computation. It unifies both time-explicit and time-implicit numerical schemes within the framework of the pre-conditioned defect-correction iteration solution procedure.
We find that time-implicit solution procedures are remarkably superior over their time-explicit counterparts in the very high $\Gamma$-regime and therefore most suitable for consistent modelling of relativistic outflows in AGNs and micro-quasars.
\end{abstract}

\begin{keywords}
methods: numerical -- hydrodynamics -- relativistic processes
\end{keywords}



\section{Introduction}
The powerful jets observed in AGNs and quasars as well as in systems containing ultra-compact objects, such as pulsars, neutron stars, magnetars or stellar black holes, have been observed to propagate with ultra-relativistic speeds \cite[see][and the references therein]{Gomez_2016}. In order to develop a deeper inside of the mechanisms underlying their initiation, their complicated internal magneto-thermal structures, the energy processes and their interaction with the surrounding media, highly robust and efficient numerical solvers are required. Here magnetic fields and radiation transfer including internal dissipative processes in multi-dimensions in the strong gravitational field regime must be taken into account \cite[see][and the references therein]{Hujeirat_2003HLCB,Brezinski_2011}. The corresponding set of equations belongs to the family of general relativistic magneto-radiative Navier--Stokes equations with an appropriate metric in the background. Hence, solving the simple ideal relativistic Euler equations here is not sufficient and therefore the stress-energy tensor should be modified considerably to include the effects of dissipation and conduction.

Although computer capacity has increased exponentially during the last three decades, carrying accurate simulations for modelling these types of plasma is still difficult and too tedious for today's computationalists.

However, in this paper we will focus on the ideal relativistic Euler equations and the basic performance of adaptive mesh refinement and parallelisation of time-explicit versus time-implicit solution procedures based on the unified solution method reported in \citep{Hujeirat_2005CoPhC,Hujeirat_2009Spectrum}.

Basically, the core of the Navier--Stokes equations is the set of Euler equations. These equations must be transferred into the finite space using an appropriate, consistent and accurate discretization strategy. Depending on the hydro-problem, one may use time-explicit or time-implicit numerical schemes to advance the numerical solution in time. Although time-explicit methods are much more popular than their time-implicit counterparts
\cite[][]{Hujeirat_2009Spectrum}, the latter continues to be superior for modelling flow configurations such as:
\begin{itemize}
  \item Quasi-stationary or time-independent
  \item Weakly-compressible or incompressible
  \item Non-ideal, diffusive and dissipative
  \item The underlying physical processes operate on much \\
  $~~~~~~~$      shorter time scales  than the hydrodynamical one\\
  $~~~~$or
  \item The density and/or temperature contrasts across the domain are relatively large.
\end{itemize}
As these are typical properties of many astrophysical fluid-flows \citep{Hujeirat_2009LowMach}, developing robust and efficient stable solvers is necessary, particularly for modelling the propagation of ultra-relativistic shock fronts; hence the aim of the present paper.

The problem of solving the relativistic hydrodynamical equations numerically has been studied for decades. The corresponding computer codes developed in the eighties were mainly based on the approach of \cite{Wilson_1972}. A Eulerian explicit finite difference scheme with monotonic transport, which turned out to be incapable of modelling relativistic flows with Lorentz factors $\Gamma > 2$ accurately. \cite{Norman_1986} developed a different method, based on a finite difference scheme including adaptive mesh refinement. They incorporated an artificial viscosity term consistent with the relativistic dynamics of non-perfect fluids. The strong coupling in the equations due to the artificial viscosity forced an implicit treatment of the equations. This was the first time of successfully capturing shock fronts with moderate Lorentz factors ($\Gamma = 3.59$). Although they intended to develop a multidimensional version of their code this never happened.

Despite the strong non-linearities underlying the transport operators, because of efficiency and simplicity reasons, most methods used in the following years were based on time-explicit solution strategies. Instead of seeking to simulate higher Lorentz factors accurately the focus was on the incorporation of additional physical processes like magnetic fields \cite[see][for further details and reviews on numerical schemes for modelling relativistic hydrodynamics]{Marti_2003,Marti_2015}. 

In section~\ref{sec:equations} we present the relativistic Euler equations to be solved numerically. We subsequently explain our numerical approach in section~\ref{sec:numerical_methods}, present our results in section~\ref{sec:results} and draw our conclusions in section~\ref{sec:conclusions}.
\section{The governing hydrodynamical Equations} \label{sec:equations}

Astrophysical jets are considered to form in the vicinity of the surfaces of central accreting objects, where the curvature of spacetime is significant and where magnetic fields in combination with radiation fields set approximately 5--10 per cent of the inflowing matter into gravitationally unbound outflowing plasmas \cite[see][and the references therein]{Hujeirat_2003HLCB}.
At a certain distance from the central object and under the effect of internal and external magnetic fields, these outflows start collimating into jets, whose plasmas set to propagate with relativistic speeds. In this regime, the spacetime is safely flat and the outflow-dimensions may be reduced into just one-dimension if transverse motions or generation of turbulence are irrelevant.\\

Under these circumstances the relativistic Euler equations become the concerned ones and they read as follows:\\
\bit
\item The continuity equation, which describes the evolution of the relativistic density $D$:
\begin{equation} \label{eq:continuity}
\frac{\partial D}{\partial t} + \frac{\partial (DV^x)}{\partial x} = 0 \, .
\end{equation}
\item The evolution of the 4-momentum equation $M_x,$ subject to pressure $P$ can be written as:
\begin{equation} \label{eq:momentum}
\frac{\partial M_x}{\partial t} + \frac{\partial (M_x V^x)}{\partial x} + \frac{\partial P}{\partial x} = 0 \, .
\end{equation}
\item The equation which describes the evolution of the internal energy density ${\cal E}^d$:
\begin{equation} \label{eq:energy}
\frac{\partial {\cal E}^d}{\partial t} + \frac{\partial ({\cal E}^d V^x)}{\partial x} + (\gamma - 1) \frac{{\cal E}^d}{u^t} \left(\frac{\partial u^t}{\partial t} + \frac{\partial(u^t V^x)}{\partial x}\right) = 0 \, ,
\end{equation}
where $\gamma$ denotes the adiabatic index and $u^t$ is the time component of the four-velocity. To close the system
of equations, the plasma is assumed to be governed by the ideal equation of state:
 \begin{equation}\label{eq:P}
P = (\gamma -1) \frac{{\cal E}^d}{u^t}.
\end{equation}
\eit
The primitive variables are extracted from the conservative variables $D$, $M_x$ and ${\cal E}^d$ as follows.
The time-component of the four-velocity is computed, using the following relation:
\begin{equation} \label{eq:ut}
u^t= \frac{\sqrt{\left(D + \gamma \, {\cal E}^d\right)^2 + M^2_x}}{D + \gamma {\cal E}^d} \, .
\end{equation}
Knowing $u^t$ from Eq.~\eqref{eq:ut}, the density of the fluid can be computed according to:
\begin{equation}\label{eq:rho}
\rho = \frac{D}{u^t}
\end{equation}
and the transport velocity of the fluid can be written as:
\begin{equation}\label{eq:Vx}
V^x = \frac{M_x}{u^t (D + \gamma \, {\cal E}^d)} \, .
\end{equation}
Additionally, the temperature can be expressed as:
\begin{equation}
T = (\gamma-1) \frac{{\cal E}^d}{D} = \frac{P}{\rho} \, .
\end{equation}
Furthermore, it should be mentioned that the adiabatic index depends on temperature for a monoatomic relativistic gas \citep{Lightman_1975,Thompson_1985}, but for the sake of simplicity, we neglect its variability, by setting $\gamma = 5/3$ in the present paper.
\section{Numerical Methods: The unified approach} \label{sec:numerical_methods}
The numerical method employed here is a simplified version of the unified approach presented by Hujeirat \cite[see][and the references therein]{Hujeirat_2005}, in which explicit methods show-up as a very special case of the preconditioned 
defect-correction iteration procedure.
To clarify the idea: Assume we are given a set of equations written in the following vector form:
\beq
 \DD{\D \mathbf{q}}{\D t} + L_q\mathbf{q}= \mathbf{b}_q \Leftrightarrow 
 R_q = \DD{\D \mathbf{q}}{\D t} + L_q\mathbf{q}- \mathbf{b}_q
 \eeq
 where $L_q$ represents the family of first-order, second-order differential operators or a combination of both, $\mathbf{b}_q$ is a vector of constant values and $R_q$ is the residual.
 Linearising the set of equations, the resulting set of linear equations may be organised in the matrix form:
 $ \mathbf{A} \, \tilde{\mathbf{q}} = \mathbf{b}, $ where $\mathbf{A}= \partial R_q/\partial \mathbf{q}$. 
 Depending on the strength of non-linearity $\mathbf{q}$ may differ significantly from $\tilde{\mathbf{q}},$ thereby giving rise to
 $\delta \mathbf{q} = \mathbf{q} - \tilde{\mathbf{q}} \neq 0.$ Moreover, in most cases, $\mathbf{A}$ may not be sparse and most likely difficult to invert. Hence, instead of inverting
 $\mathbf{A}$ one may try to construct a simplified matrix $\tilde{\mathbf{A}}$, that must fulfil the following conditions:
\begin{itemize}
 \item $\tilde{\mathbf{A}}$ should be easy to invert
 \item $\tilde{\mathbf{A}}$ and $\mathbf{{A} }$ are similar, i.e. both matrices share the same spectral properties. For completeness: two matrices
 $\mathbf{A}$ and $\mathbf{B} $ are said to be similar, if there exists a matrix $\mathbf{P}$ such that $ {\mathbf{A} = \mathbf{P}^{-1}\mathbf{B}\,\mathbf{P}}$, where the columns of $\mathbf{P}$ consist
 of the eigenvectors of $\mathbf{A}$. In this case, the eigenvalues of $\mathbf{A}$ and $\mathbf{B}$ are the same: hence the meaning of sharing the same spectral space.
\end{itemize}
Noting that the solution of the linear system
${\tilde{\mathbf{A}}} ~\mathbf{\tilde{\tilde{q}}} = \mathbf{b}$ may differ considerably from the solution of the original
matrix equation ${\mathbf{A}} \tilde{\mathbf{q}} = \mathbf{b}$, a constraining mechanism is required to ensure consistency of the mathematical formulation with the
original physical problem. This can be done by employing the defect-correction strategy, namely instead of solving
$\mathbf{{A} }\tilde{\mathbf{q}} = \mathbf{b}$, we solve:
\beq
 \tilde{\mathbf{A}} \mathbf{\mu} = \mathbf{d} \doteq \mathbf{b} - \mathbf{A }{\mathbf{q}^*},
\eeq
where $\mathbf{\mu}$ and ${\mathbf{q}^*}$ are respectively the correction and the transient solution that, after several iterations, should converge to the true solution of the non-linear system.

Consequently, the closer the preconditioner $\tilde{\mathbf{A}}$ is to the original matrix $\mathbf{{A} }$ the smaller is the number of iterations needed for $\mathbf{\mu}$ to converge to zero, or equivalently, to solving the set of linear equations.\\
Noting that the identity matrix $\mathbf{I}$ is the easiest one to invert, one may ask whether the identity $\mathbf{I}$ could be reliably used as a preconditioner for $\mathbf{{A} }$?\\
\begin{figure}
\centering {\hspace*{-0.95cm}
\includegraphics*[angle=-0, width=9.15cm]{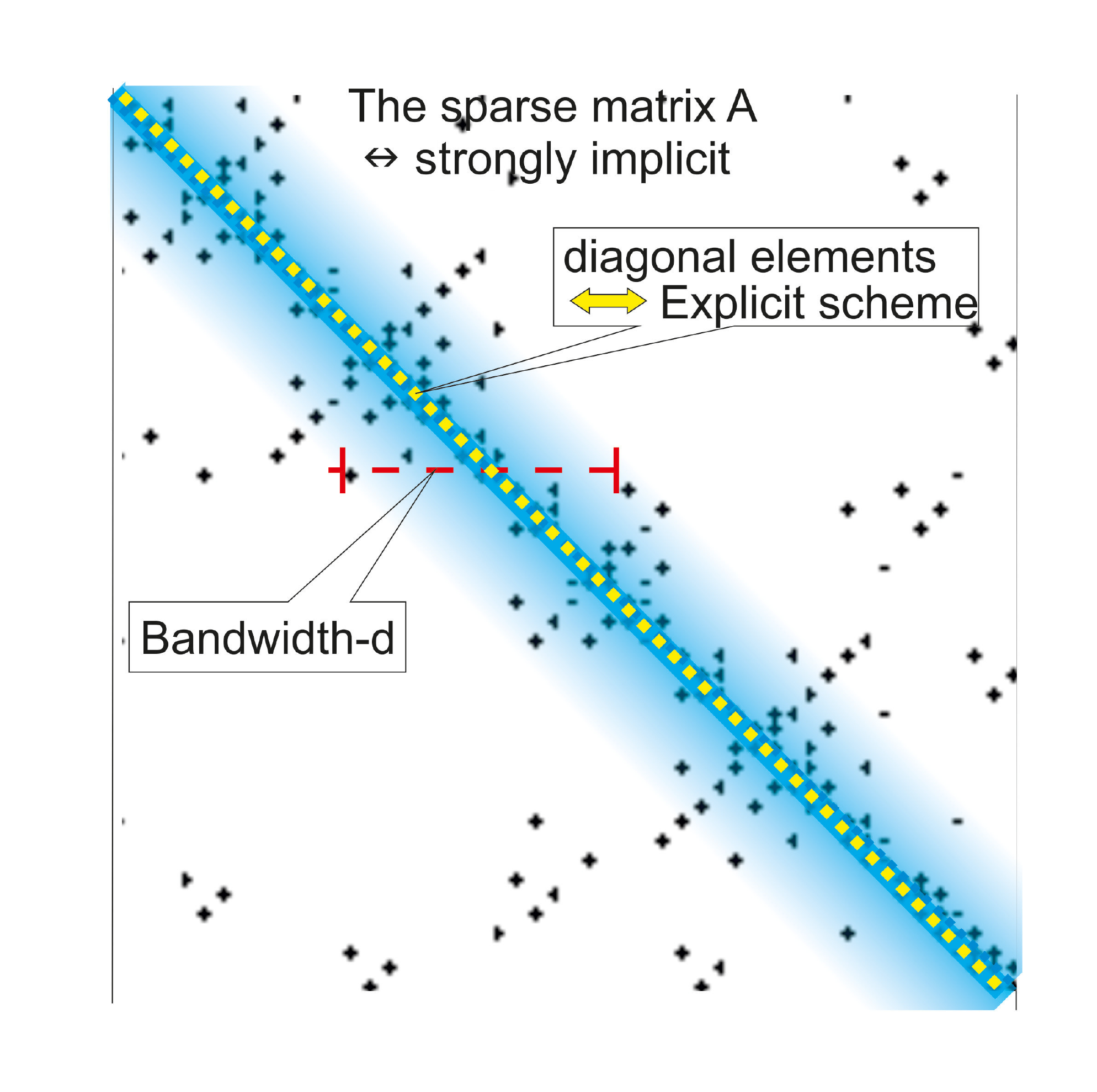}
}
\caption{ \small A schematic description of the Jacobian, $J$, corresponding to the linearised system. $J$ may be reduced into a 
band matrix with bandwidth $d$. The computational costs scale 
as: $CC \sim N \cdot d^2,$ where $N$ denotes the number of grid points times the number of equations. Explicit methods correspond to $N=1$, where the $CC$ attains a minimum: hence the origin of their unrivalled 
efficiency, though have the lowest robustness.
}\label{fig:SMatrix}
\end{figure}
If matrix $\mathbf{{A} }$ corresponds to time-dependent fluid flows or to plasma motions, then $\mathbf{{A} }$ may be decomposed into the two matrices:
 \beq
 \mathbf{{A} } = \left(\DD{1}{\delta t} \mathbf{I} + \mathbf{R}\right).
 \eeq
 Or equivalently:
 \beq
 \mathbf{{A} } = \left(\DD{1}{\delta t}\right) \, \mathbf{I} \, ( \mathbf{I} + \delta t \mathbf{R})
 \Longrightarrow \mathbf{{A}}^{-1} = ( \mathbf{I} + \delta t \mathbf{R})^{-1} \left(\DD{\mathbf{I}}{\delta t} \right)^{-1} \, .
 \eeq
The matrices $ \mathbf{{A} }$ and $\DD{1}{\delta t} \mathbf{I}$ may share the same spectral space (i.e. they have the same eigenvalues),
 if the norm of $\delta t ~ ||\mathbf{R}|| \ll 1$. A relevant measure here would be the maximum norm:
 $|| \mathbf{R}||_\mathrm{max} = \sum_i |a_i|,$ where $a_i$ are the elements of $|| \mathbf{R}||$. This is, by the way, the necessary condition for inverting a matrix stably \citep{Hackbusch_1994}. \\
 On the other hand, the matrix $ (\mathbf{I} + \delta t \mathbf{R})^{-1}$ may be expanded into the infinite power series as follows:
 \beq
 (\mathbf{I} + \delta t \mathbf{R})^{-1} = \sum^\infty_{n=0} (-1)^{n} {\delta t}^{n+1} \mathbf{R}^n \, .
 \eeq
 This power series converges, if the ${\delta t} \, ||\mathbf{R}|| < 1$. Applying this analysis to the 1D Euler or Navier--Stokes equations, then the entries $a_i$ of $\mathbf{R}$ must be of type $u_i/\Delta x$, $\nu_i/\Delta x^2$ or some combination of both. $u_i, \nu_i \mbox{ and } \Delta x $ correspond to the fluid velocity in the finite space, viscosity coefficient and grid spacing at a finite distance $x_i$, respectively. Putting terms together, the condition for stably inverting the coefficient matrix or for sharing the same spectral space is equivalent to require:
 \beq
 \delta t \left( \DD{|u_i|}{\Delta x}+ \DD{\nu_i}{\Delta x^2}\right) \ll 1 \, ,
 \eeq
 which is equivalent to the well-known Courant--Friedrichs--Lewy condition $\mathcal{C}_\mathrm{CFL} \doteq \delta t \left( \DD{|u_i|}{\Delta x}+ \DD{\nu_i}{\Delta x^2}\right) <1.$ 
 
A time-explicit method is a very special case of the preconditioned defect-correction iteration procedure in which the identity matrix $\mathbf{I}$
 is used as a preconditioner and where just one iteration per time step is performed only. However, using this strategy requires that the elements of $\delta t~ \mathbf{R}$ must be negligibly small compared to the diagonal elements of $ \mathbf{I}$, which is, in the case of fluid equations, equivalent to the requirement: $\mathcal{C}_\mathrm{CFL} <1$. \\
 The stability condition, $\mathcal{C}_\mathrm{CFL} <1$, appears to be equivalent to requiring $\tilde{\mathbf{A}}$ be diagonally dominant and that this can be safely fulfilled if the time-step size is sufficiently small.

 This implies that there must be a sequence of preconditionings: $ \{\mathbf{A},...,\mathbf{A}_j,... \mathbf{I} \},$ in which the degree of implicitness decreases gradually from the strongly implicit case: $ \tilde{\mathbf{A}} = \mathbf{A}$ down to the pure explicit case: $ \tilde{\mathbf{A}} = \mathbf{I}$.
 The above sequence of matrices differ from each other through their
 bandwidth $d$. As the computational costs scales as $CC\sim N \cdot d^2$,
 where $N$ is the number of grid points times the number of equations,
 we conclude that the smaller the bandwidth $d$ is, or equivalently, the more off-diagonal entries are neglected, the smaller is the $CC$ and therefore the weaker the implicitness of the matrix would be (see Fig.~\ref{fig:SMatrix}).
 
 Depending on the physical problem in hand, taking $ \tilde{\mathbf{A}} = \mathbf{A}$ can be used for modelling quasi-stationary, weakly compressible and highly dissipative flows with complicated chemical and radiative processes, whereas $\tilde{\mathbf{A}} = \mathbf{I}$ is optimally used for modelling strongly compressible, almost ideal (non-dissipative) and strongly time-dependent plasma motions, such as turbulent generation and/or propagation of shock fronts.\\

 The strongly implicit methods used in the former case must be highly robust, though the associated computational costs may become prohibitively expensive,
 as the inversion procedure must take the whole elements of the Jacobian into account, thereby damaging the sparsity of the matrix through the fill-in effect. However, one may circumvent this difficulty by using Krylov subspace iteration methods, where advantages of the sparsity of the matrix 
 are almost maintained.
 On the other hand, the efficiency of explicit methods is unrivalled as the computational costs per time-step are lowest, though time-marching is extraordinarily slow and requires a very large number of time steps to cover relevant time scales. \\

\subsection{Discretization method}
\begin{figure}
\centering {\hspace*{0.75cm}
\includegraphics*[angle=-0, width=7.15cm]{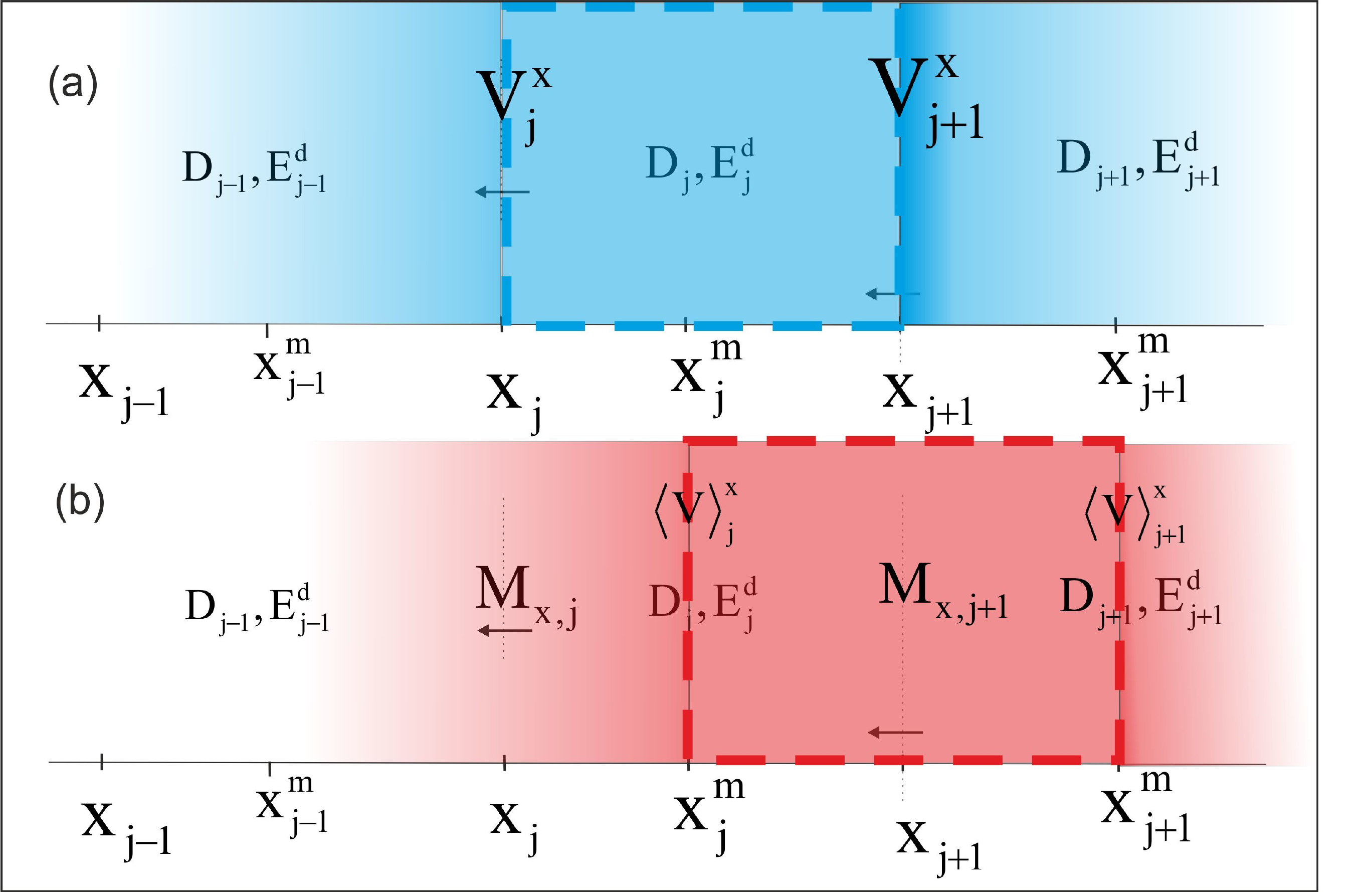}
}
\caption{\small A schematic representation of the finite volume method applied to scalars (a) and the shifted finite volume cell (staggered cell) for computing the momentum (b). In the former case, scalars, such as density $\rho$, relativistic density $D$, density of internal energy ${{\cal E}}^d$ and the time component of the four-velocity $u^t$ are defined at the cell centres whereas velocities and fluxes, such as contravariant transport velocity $V^x$ and the covariant momentum flux $M_x$ are defined at the cell surface, in accordance with divergence theorem. 
} \label{fig:grid}
\end{figure}

To solve the relativistic Euler equations, we use the finite volume formulation to ensure local conservation of mass, internal energy and momentum. The equations are discretized using one-dimensional finite volume cells. Scalar quantities, such as $D$, ${\cal E}^d$, $\rho$ and $P$ are defined at cell-centres, whereas the corresponding fluxes are defined at cell-surfaces. For evaluating
the momentum, the staggered grid discretization strategy is employed (see Fig.~\ref{fig:grid}).

Within each time step, several additional iterations are performed. Firstly, the equations~\eqref{eq:continuity}--\eqref{eq:energy} are solved for the main variables, then followed by computing the primitive variables according to equations~\eqref{eq:ut}--\eqref{eq:Vx}. This information is used in the following iteration and so on until a stop criterion is fulfilled.

For simplicity and test purposes we split the unified solver into a purely time-explicit and time-implicit solution procedure.
As stability requirements of explicit schemes require the time step size to be extremely small, a first-order temporal accuracy would be sufficient generally. However, this would fail for time-implicit schemes as the time step size is here theoretically unlimited, but limited due to accuracy reasons. Indeed in the present calculations, the underlying phenomenon is highly time-dependent and therefore we limited the $\mathcal{C}_\mathrm{CFL} \leq 1/2$ for accuracy and stability reasons, which is 50 times larger than the maximum time-step used in the time-explicit version of the solver. Technically, a specific value of $s = \Delta t / \Delta x$, that fulfils the stability and accuracy requirements is chosen, from which the time step $\Delta t$ is determined.

\subsection{The time-explicit procedure}
Our unified numerical approach may be reduced into a time-explicit one by setting the preconditioning $ \tilde{\mathbf{A}} = (\DD{1}{\delta t})~\mathbf{I}$.
In this case, the matrix equation corresponding to equations~\eqref{eq:continuity}--\eqref{eq:energy} in the finite space read as follows:
\beq
\tilde{\mathbf{A}} \, \mathbf{\mu} = \mathbf{d} \Leftrightarrow(\DD{1}{\delta t})~\mathbf{I} \, \mathbf{\mu} = \mathbf{d} \, ,
\eeq
where $ \mathbf{\mu} = \mathbf{q}^{\star}_\mathrm{new}-\mathbf{q}^{\star}_\mathrm{old}$ and $\mathbf{q}^{\star}_\mathrm{new}$ is the intermediate solution which is, in the absence of local iteration, identical to the sought solution
$\mathbf{q}^{n+1}$. $\mathbf{q}^\star_\mathrm{old}$ here corresponds to $\mathbf{q}^n$. In this case the matrix equation can be reduced to component-wise equations:
\beq
\DD{\mathbf{q}^{n+1}-\mathbf{q}^n}{\Delta t} = \mathbf{d}^n \Leftrightarrow \mathbf{q}^{n+1} = \mathbf{q}^n + \Delta t \cdot \mathbf{d}^n \, .
\eeq
As $\mathbf{q} = \left(D, \, M_x, \, {\cal E}^d\right)^\mathsf{T}$ the equations read:
\begin{flalign} \label{eq:continuity-dis}
 D^{n+1}_j =& \, D^n_j + \frac{\Delta t}{\Delta x} \left(\vec{D}^{n}_{j-1/2} V^{x,n}_{j-1/2} - \vec{D}^{n}_{j+1/2} V^{x,n}_{j+1/2}\right) &&\nonumber\\
 &+ \Delta t \, Q_{\mathrm{diff},D} \, ,&&
\end{flalign}
\begin{flalign} \label{eq:momentum-dis}
M^{n+1}_{x,j-1/2} =& \, M^n_{x,j-1/2} + \frac{\Delta t}{\Delta x} \left(\vec{M}^{n}_{x,j-1} {\langle V^x \rangle}^{n}_{j-1} \right. &&\nonumber\\
&- \left. \vec{M}^{n}_{x,j} {\langle V^x \rangle}^{n}_{j} - \Delta P_{j-1/2}^n \right) &&\nonumber\\
&+ \Delta t \, Q_{\mathrm{diff},M_x} \, ,&&
\end{flalign}
\begin{flalign} \label{eq:energy-dis}
{\cal E}^{d,n+1}_{j} =& \, {\cal E}^{d,n}_{j} + \frac{\Delta t}{\Delta x} \left(\vec{{\cal E}}^{d,n}_{j-1/2} V^{x,n}_{j-1/2} - \vec{{\cal E}}^{d,n}_{j+1/2} V^{x,n}_{j+1/2} \right) &&\nonumber\\
&- \, P^n_j \left( u^{t,n+1}_j -u^{t,n}_j + \frac{\Delta t}{\Delta x} \, \Delta(u^t \, V^x)^n_j \right) &&\nonumber\\
&+ \Delta t \, Q_{\mathrm{art},{\cal E}^d} + \Delta t \, Q_{\mathrm{diff},{\cal E}_d} \, .&&
\end{flalign}
$\vec{D}$, $\vec{M_x}$, $\vec{{\cal E}}^d$ are the upwind values of $D$, $M_x$, ${\cal E}^d$ respectively. Hence the multiplication with the transport velocity $V^x$ returns the flux through a cell interface. Note, the velocity $\langle V^x_j \rangle$ is not evaluated at the cell centres, but at the interfaces of the staggered gird, which is only the same when a uniform grid is used. Further details can be found in section~\ref{sec:update_primvar}. The transport operators are described in detail in section~\ref{subsec:trans_op}, especially, the differences $\Delta P$ and $\Delta(u^t \, V^x)$ are given in equations~\eqref{eq:dp_fo}--\eqref{eq:duV_so}.
The energy equation contains a time-derivative of the general Lorentz factor $u^t$, which implies that this equation can be viewed as an evolutionary equation for both ${\cal E}^{d}$ and $u^t$, or alternatively as an additional algebraic constraint. However, the equation must be solved for each time step iteratively, using the Newton--Raphson method.
The $Q$-terms originally not belonging to the Euler equations are described in sections~\ref{sec:art_visc} and \ref{sec:overshooting}.

\subsection{The time-implicit procedure}
Based on the unified solution method, the matrix equation $\tilde{\mathbf{A}} \mathbf{\mu} = \mathbf{d}$ is now solved using the following strategy:
\begin{itemize}
\item The preconditioning $\tilde{\mathbf{A}}$ is constructed using a first-order discretization method in space and time. This is necessary to ensure strong diagonal dominance of the matrix.
\item The defect $\mathbf{d}$ is evaluated at the new time level, i.e. $\mathbf{d} = \mathbf{d}(D,M_x,E^d)^{*},$ where highly accurate spatial and temporal accuracy schemes are used. Note that the intermediate value of $\mathbf{d}^{*}$ may differ from $\mathbf{d}^{n+1}$ due to the non-linearities characterising the transport operators. This deviation may be reduced through performing iterations within a time step.
\end{itemize}

For achieving second-order temporal accuracy we discretise derivatives as described as follows. The advantage of the implicit scheme is that we can make use of values from $n$ and $n+1$, although we are computing values at time $n+1$. 

For the continuity equation, the formulation is the same as for the explicit scheme (Eq.~\eqref{eq:continuity-dis}), but we use a different formulation of the fluxes. This formulation computes fluxes at $n+1/2$, which is time-implicit. The following flux formulation is used in all three relativistic Euler equations.
\begin{equation} \label{eq:flux_imp}
f_{j+1/2}^{n+1/2} = \frac{1}{\Delta t} \int_{0}^{\Delta t} q(t) \cdot u(t) \, dt \,.
\end{equation}
The computation of the flux $f_{i+1/2}^{n+1/2}$ is based on the assumption, that $q$ and the corresponding transport velocity $u$ are linear functions within a time step.
\begin{equation}
q(t) = q_{j+1/2}^n + t \cdot \frac{q_{j+1/2}^{n+1}-q_{j+1/2}^{n}}{\Delta t}.
\end{equation}
\begin{equation}
u(t) = u_{j+1/2}^n + t \cdot \frac{u_{j+1/2}^{n+1}-u_{j+1/2}^{n}}{\Delta t}.
\end{equation}
Note, $q$ here corresponds to its value at the cell interface, which is obtained using the subgrid model of the flux limiter described in section~\ref{subsec:trans_op}.
Using the flux $f^{n+1/2}_{j-1/2}$ for $\vec{D}^{n+1/2}_{j-1/2} V^{x,n+1/2}_{j-1/2}$ the implicit continuity equation can be written as:
\begin{flalign} \label{eq:continuity-dis_imp}
 D^{n+1}_j &= D^n_j + \frac{\Delta t}{\Delta x} \bigg[ \left(\vec{D}^{n+1/2}_{j-1/2} V^{x,n+1/2}_{j-1/2} - \vec{D}^{n+1/2}_{j+1/2} V^{x,n+1/2}_{j+1/2}\right) &&\nonumber\\
 &+ \Delta x \, Q_{\mathrm{diff},D} \bigg] \, .&&
\end{flalign}
For the momentum equation we compute the spatial derivative $\Delta P^{n+1/2}_{j-1/2} / \Delta x$. This derivative is computed using values at $n+1/2$, which are obtained by taking the mean of the pressure:
\begin{equation}
P^{n+1/2}_j = \frac{P^n_j + P^{n+1}_j}{2} \,.
\end{equation}
The formulation of the derivative is given by Eq.~\eqref{eq:dp_fo} or \eqref{eq:dp_so}, depending on the desired spatial accuracy. The implicit momentum equation can be written as:
\begin{flalign} \label{eq:momentum-dis_imp}
M^{n+1}_{x,j-1/2} &= M^n_{x,j-1/2} + \frac{\Delta t}{\Delta x} \bigg[ \left(\vec{M}^{n+1/2}_{x,j-1} {\langle V^x \rangle}^{n+1/2}_{j-1} \right. &&\nonumber\\
&- \left. \vec{M}^{n+1/2}_{x,j} {\langle V^x \rangle}^{n+1/2}_{j} - \Delta P_{j-1/2}^{n+1/2} \right) &&\nonumber\\
&+ \Delta x \, Q_{\mathrm{diff},M_x} \bigg] \, .&&
\end{flalign}
The energy equation is discretized as follows:
\begin{flalign} \label{eq:energy-dis_imp}
{\cal E}^{d,n+1}_{j} &= \, {\cal E}^{d,n}_{j} + \frac{\Delta t}{\Delta x} \, \Bigg[\left(\vec{\cal E}^{d,n+1/2}_{j-1/2} V^{x,n+1/2}_{j-1/2} - \vec{\cal E}^{d,n+1/2}_{j+1/2} V^{x,n+1/2}_{j+1/2} \right) &&\nonumber\\
&- \frac{P^n_j + P^{n+1}_j}{2} \, \left( \frac{u^{t,n+1}_j - u^{t,n}_j}{\Delta t} \, \Delta x + \Delta (u^t V^x)^{n+1/2}_j \right) &&\nonumber\\
&+ \Delta x \, Q_{\mathrm{art},{\cal E}^d} + \Delta x \, Q_{\mathrm{diff},{\cal E}_d} \Bigg]\, .&&
\end{flalign}
Here we use:
\begin{equation}
\Delta (u^t V^x)^{n+1/2}_j = \frac{\Delta (u^t V^x)^n_j + \Delta (u^t V^x)^{n+1}_j }{2} \, .
\end{equation}
The spatial derivative of $u^t V^x$ is computed according to Eq.~\eqref{eq:smearing}. 

The defect-correction iteration procedure is used to solve the differential equations implicitly.
This implies that Eq.~\eqref{eq:def-cor} is solved 
for $\mathbf{\mu}$ iteratively to recover the second-order temporal accuracy.
\begin{equation} \label{eq:def-cor}
\tilde{\mathbf{A}} \, \mathbf{\mu} = \mathbf{d},
\end{equation}
where $\mathbf{d}$ is the defect, which is defined as follows:

\begin{equation} \label{eq:defect}
d_j = \frac{q^{*}_j-q^{n}_j}{\Delta t} - \frac{1}{\Delta x} \left[ L_j \left(q^{n}, q^{*} \right) \right] \, .
\end{equation}
Here $L_j$ corresponds to the term in square brackets of equations~\eqref{eq:continuity-dis_imp}, \eqref{eq:momentum-dis_imp} and \eqref{eq:energy-dis_imp}.
Equation~\eqref{eq:def-cor} is constructed and applied for each of the three relativistic Euler equations. ${}^*$ denotes the intermediate value of the corresponding variable within each time step. The value returned by the operator $L_j$ depends on both, the last time step $n$ and the intermediate value ${}^*$.
The stopping criterion of the iteration procedure is based on the summation over the defect of all cells. The procedure ends, once
\begin{equation} \label{eq:stop_crit}
\sum_j d_j(D) + \sum_j d_j(M_x) + \sum_j d_j({\cal E}^d) \leq \epsilon \, .
\end{equation}
The matrix $\tilde{\mathbf{A}}$ of Eq.~\eqref{eq:def-cor} can be written as:
\begin{flalign}
\qquad \tilde{\mathbf{A}}= \left(\begin{matrix}
\ddots & \ddots & 0 & 0 &0\\
\frac{\partial R_{j-1}}{\partial q^*_{j-2}} & \frac{\partial R_{j-1}}{\partial q^*_{j-1}} & \frac{\partial R_{j-1}}{\partial q^*_{j}} & 0 & 0\\
0&\frac{\partial R_j}{\partial q^*_{j-1}} & \frac{\partial R_j}{\partial q^*_{j}} & \frac{\partial R_j}{\partial q^*_{j+1}}&0\\
0&0& \frac{\partial R_{j+1}}{\partial q^*_{j}} & \frac{\partial R_{j+1}}{\partial q^*_{j+1}} & \frac{\partial R_{j+1}}{\partial q^*_{j+2}}\\
0 & 0 &0 & \ddots & \ddots\\
\end{matrix}\right) \, .
\end{flalign}
Here we use the residual $R_j$ instead of $d_j$, which we define as
\begin{equation} \label{eq:residual}
R_j = \frac{q^{*}_j-q^{n}_j}{\Delta t} - \frac{1}{\Delta x} \tilde{L}_j \, ,
\end{equation}
where $\tilde{L}_j$ is a first-order (spatial accurate) version of $L_j$. For which we usually do not take artificial viscosity or diffusion terms ($Q$-terms) into account.

The iteration procedure solves the continuity, momentum and energy equation in parallel, i.e. the matrices for all three are inverted in parallel and then the main variables are corrected according to Eq.~\eqref{eq:main-correction}.
\begin{equation} \label{eq:main-correction}
\mathbf{q}^{\,*}_\mathrm{new} = \mathbf{q}^{\,*}_\mathrm{old} + \mathbf{\mu}
\end{equation}
Next, the primitive variables are updated to compute $\mathbf{d}$ and then the next iteration of the procedure starts, but only if $\mathbf{d}$ isn't very small, according to Eq.~\eqref{eq:stop_crit}.

\subsection{Transport operators and other derivatives} \label{subsec:trans_op}
The transport operators are discretized using flux-limiters. Accordingly, these techniques are expected to provide accurate values of the fluxes in critical and dynamically active regions and should enhance the spatial accuracy up to second order in most cases.
They are widely spread tools for following shock fronts. Fluxes of the time-explicit scheme are calculated according to Eq.~\eqref{eq:fl_f}, which makes use of three further definitions, Eq.~\eqref{eq:fl_T} and Eq.~\eqref{eq:fl_r} and the flux limiter itself, for instance Eq.~\eqref{eq:sweby}. We use $u$ to denote the transport velocity in respect to the main variable.
\begin{equation} \label{eq:fl_T}
\Theta_{j-1/2} = \left\lbrace \begin{array}{ccc}
+1 &\textrm{if}& u_{j-1/2} > 0 \\
-1 &\textrm{if}& u_{j-1/2} \le 0
\end{array} \right.
\end{equation}
\begin{equation} \label{eq:fl_r}
r_{j-1/2} = \left\lbrace \begin{array}{ccc}
\frac{q_{j-1}-q_{j-2}}{q_{j}-q_{j-1}} &\textrm{if}& u_{j-1/2} > 0 \\
\frac{q_{j+1}-q_{j}}{q_{j}-q_{j-1}} &\textrm{if}& u_{j-1/2} \le 0
\end{array} \right.
\end{equation}

\begin{flalign} \label{eq:fl_f}
f_{j-1/2} &= \frac{1}{2}\,u_{j-1/2} \cdot \left[(1+\Theta_{j-1/2})\,q_{j-1} + (1-\Theta_{j-1/2})\,q_{j}\right] &&\nonumber\\
&+ \frac{1}{2} |u_{j-1/2}| \left(1-\frac{u_{j-1/2} \, \Delta t}{\Delta x}\right) \cdot \Phi(r_{j-1/2}) \, (q_{j}-q_{j-1}) &&
\end{flalign}
Note, that the time-implicit fluxes are computed according to Eq.~\eqref{eq:flux_imp} using the same linear subgrid model as expressed in Eq.~\eqref{eq:fl_f}.
We implemented several flux limiters, the Sweby-limiter appears to provide stable and 
relatively accurate solutions \cite[see][]{Sweby_1984}.
\begin{flalign} \label{eq:sweby}
\Phi_\mathrm{sweby}(r)\,= \,\max\left(0,\min\left(\beta \, r,1\right),\min\left(r,\beta\right)\right)\nonumber\\
\textnormal{ with } \beta \in [1,2], \qquad \lim _{{r\rightarrow \infty }}\Phi _\mathrm{sweby}(r)=\beta
\end{flalign}
However, we found the generalised minmod limiter to behave better for our purpose \citep{vanLeer_1979,Harten_1987,Kurganov_2002}.
\begin{flalign}
\qquad \Phi_{mg}(r)\,= \,\max\left(0,\min\left(\beta \, r,\frac{1+r}{2}, \beta \right)\right)\nonumber\\
\textnormal{ with } \beta \in [1,2], \qquad \lim _{{r\rightarrow \infty }}\Phi _{mg}(r)=2
\end{flalign}
The parameter $\beta$ can be chosen in the given range, it controls the diffusivity. In our study we used $\beta=1.5$.

The transport operators of the continuity and energy equation can be directly expressed as written above (equations~\eqref{eq:fl_T}--\eqref{eq:fl_f}). For the momentum equation, it is slightly different, because we store the $M_x$-values at the cell centres of the staggered grid. Nevertheless, we use the equations above but evaluated for the staggered grid. Therefore we use the velocity $\langle V^{x} \rangle$, which is defined at the interfaces of the staggered cells. It is computed together with the other primitive variables. Further information is given in section~\ref{sec:update_primvar}.

Besides the transport operators, we have to specify the remaining spatial derivatives $\Delta P / \Delta x$ and $\Delta(u^t \, V^x) / \Delta x$. The first-order version can be written as:
\begin{equation} \label{eq:dp_fo}
\Delta P_{j-1/2}^n = P^n_{j} - P^n_{j-1}
\end{equation}
\begin{flalign} \label{eq::duV_fo}
\Delta(u^t \, V^x)^n_{j-1/2} &= V_j^{x,n} \, \left(u_{j+1}^{t,n} - u_{j}^{t,n} \right) && \nonumber\\
&+ u^{t,n}_j \, \left( V^{x,n}_{j+1/2} - V^{x,n}_{j-1/2} \right) &&
\end{flalign}
To achieve a higher spatial accuracy we use:
\begin{equation} \label{eq:dp_so}
\Delta P_{j-1/2}^n = \frac{P^n_{j-2} - 27 \, P^n_{j-1} + 27 P^n_{j} - P^n_{j+1}}{24} \, ,
\end{equation}

\begin{flalign} \label{eq:duV_so}
\Delta(u^t \, V^x)^n_j &= V^{x,n}_{j} \, \Delta u^{t,n}_j \, + \, u^{t,n}_j \, \Delta V^{x,n}_j &&\nonumber\\
&= V_{j}^{x,n} \, \frac{u_{j-3/2}^{t,n} - 27\,u_{j-1/2}^{t,n} + 27\,u_{j+1/2}^{t,n} - u_{j+3/2}^{t,n}}{24} &&\nonumber\\
&+ \, u^{t,n}_j \, \frac{V^{x,n}_{j-3/2} - 27\,V^{x,n}_{j-1/2} + 27\,V^{x,n}_{j+1/2} - V^{x,n}_{j+3/2}}{24} \, . &&
\end{flalign}
In practice we use a more general formulation for $\Delta(u^t \, V^x)^n_j$ (see Eq.~\eqref{eq:smearing}). This is because of numerical problems we encountered, for details see section~\ref{sec:overshooting}.

\subsection{Update of primitive variables} \label{sec:update_primvar}
The update of the primitive variables is not as simple as it may seem according to equations~\eqref{eq:P}--\eqref{eq:Vx}. We need to compute some primitive variables not only for the normal cells but also for the staggered cells and the interfaces of the normal cells, e.g. the transport velocity. Note, the staggered cells and the interfaces of the normal cells are only the same when considering an equally spaced grid. To compute the primitive variables where needed, first, the main variables are evaluated at the corresponding location. This is done by using the linear subgrid model of the used flux limiter. Then the primitive variables are computed from these values.

\begin{figure}
\includegraphics[width=\columnwidth]{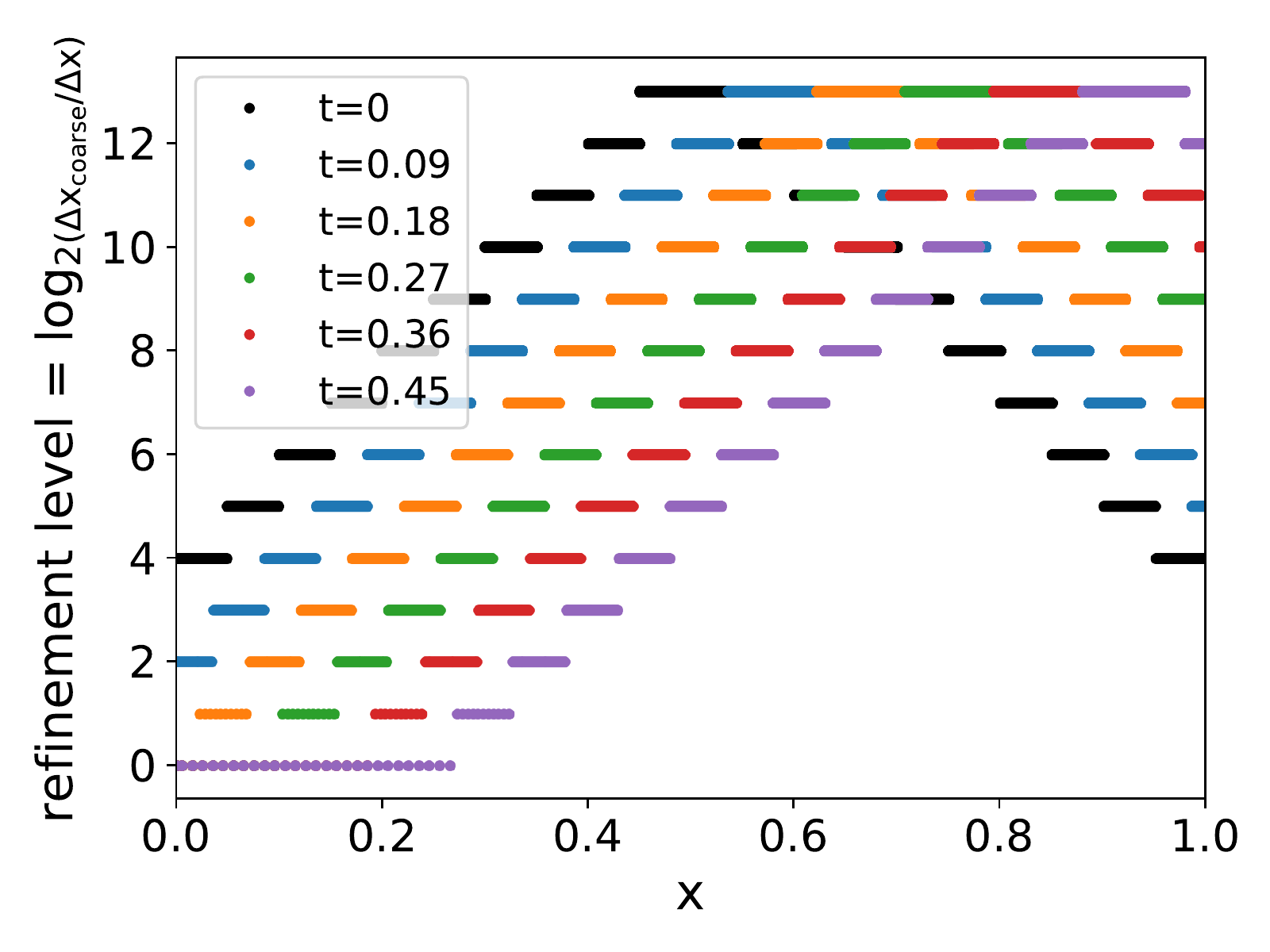}
\caption{The spatial distribution of refinement levels at different elapsed times shown with different colours (see also section~\ref{sec:results}). A certain refinement level reflects the number of halvings of a coarse grid cell, $N =N(x,t),$ during the course of the calculations. Obviously, at the shock front, where
the maximum spatial resolution is required, the number of halvings reaches 13, which implies that the initial size of the cell has been
decreased by $(1/2)^{13}$. Therefore the final grid distribution is highly non-linear with an aspect ration of order $\sim 10^4$. In fact, the regions with maximum spatial variation are 8192 more highly resolved than regions with the smoothest gradients. \label{fig:refinement-level}}
\end{figure}

\subsection{Adaptive mesh refinement}
Adaptive mesh refinement is a strategy for enhancing the spatial resolution in critical regions. Only in these regions, the grid point density is increased. To refine the grid we apply
the cell-by-cell refinement strategy. For adding a grid point a cell is split into two cells of equal size and for derefinement two neighbouring cells of equal size are joined together to form a new, bigger one. As refinement criterion gradients of pressure $P$ and relativistic density $D$ are applied.
Refinement of the grid is done in four steps. The first step is about tagging cells for refinement/derefinement.
\begin{equation}
\xi(q_j) = \max\left(\frac{q_{j-1}}{q_{j}}+\frac{q_{j}}{q_{j-1}},\frac{q_{j}}{q_{j+1}}+\frac{q_{j+1}}{q_{j}}\right)
\end{equation}
A cell is tagged for refinement if $\xi(P_j)>2.2$ or $\xi(D_j)>2.2$ and if $\xi(P_j)<2.04$ and $\xi(D_j)<2.04$ the cell is tagged for derefinement. If none of both is true the cell is tagged for prohibiting derefinement in its neighbourhood.
In a second step, it is decided, which cell is refined and which is derefined. Therefore a refinement length, which is defined in physical units, is applied to the tagged cells. This means that all cells next to a cell that is tagged for prohibiting derefinement and are within the refinement length are protected from derefinement. The selection of cells for refinement works in the same way, which means that all cells within the refinement length are refined. Additionally, to prohibiting derefinement also cells beyond the refinement length can be affected, because the implementation also ensures a step-shaped refined grid as can be seen in Fig.~\ref{fig:refinement-level}.
This means that within the distance of one refinement length the refinement level can only drop ones.
The splitting and recombination of cells is done in a third step. Therefore only the main variables ($D$, $M_x$, ${\cal E}^d$) are considered. For splitting cells a linear sub-grid model is assumed, which is the same as the one used by the flux limiter applied in the transport operators. According to this, the refinement scheme is in some sense of second spatial order. In the fourth and last step, the primitive variables are computed using the new main variables from the previous step. Furthermore, we use a global time step in our scheme, so that $\Delta t$ is set by the size of the smallest cell. Consequently, we don't apply adaptive time-stepping as in other AMR codes, like the RAMSES code \cite[see ][]{Teyssier_2002,Commercon_2014}.
We don't adapt the grid each time step, rather the time between to adaptions is chosen in such a manner that the physical features of interest can't propagate outward the highest resolved region.

Furthermore, it should be mentioned that all the equations of this section were expressed for a regularly spaced grid. This is no longer true in AMR. The discretization of derivatives and the update of primitive variables depends on grid spacing and without bothering the reader with too many details we skip the exact formulation, which especially with a staggered grid becomes more complicated.

\subsection{Artificial viscosity} \label{sec:art_visc}
Most high order advection schemes do not respect monotonicity across shock fronts but are affected from under and over-shooting, thereby deviating considerably from the physical or analytical solution. While reducing the accuracy to first order in these critical regions is a default suggestion, the strong numerical solution here could affect the solution in the whole domain. A promising strategy is incorporating shock-capturing techniques, which rely on constructing an artificial viscosity operator that operates solely across shock fronts, but vanishes elsewhere. Using such a second-order viscosity operator here would maintain communications between the fluids in the pre- and post-shock regions, thereby inhibiting the formation of over and under-shooting. In the present study, we define the kinematic coefficient of the artificial viscosity:
\begin{equation}
\nu_\mathrm{art} = \left\lbrace \begin{array}{ccc}
-\alpha_\mathrm{art} \, (\Delta x)^2 \, \left(\frac{\partial V^x}{\partial x}\right) &\textrm{if}& \frac{\partial V^x}{\partial x} < 0 \vspace{0.1cm}\\
0 &\textrm{if}& \frac{\partial V^x}{\partial x} \ge 0, 
\end{array} \right. \, .
\end{equation}
where $\alpha_\mathrm{art}$ is a constant coefficient, which is chosen to reproduce the exact solution of the test problems we study. Despite the idea of reducing over and undershooting, the artificial viscosity term is for our study most important for reproducing the correct Lorentz factors but only for the time-explicit scheme.
The viscosity coefficient is incorporated in the second-order diffusion operator:
\begin{equation} \label{eq:artvisc_ed}
Q_{\mathrm{art},{\cal E}^d} = \eta_\mathrm{art}
\left(\frac{\partial V^x}{\partial x}\right)^2,
\end{equation}
where $\eta_\mathrm{art} = \nu_\mathrm{art} D$ is the dynamical coefficient of the artificial viscosity.\\
Obviously, the effect of $Q_{\mathrm{art},{\cal E}^d}$ is significant only across the shock fronts,
where the velocity gradient is large but decays exponentially in smooth regions, where
$\frac{\partial V^x}{\partial x}$ is small.
This operator is applied to the internal energy equation only.

\begin{figure}
\includegraphics[width=\columnwidth]{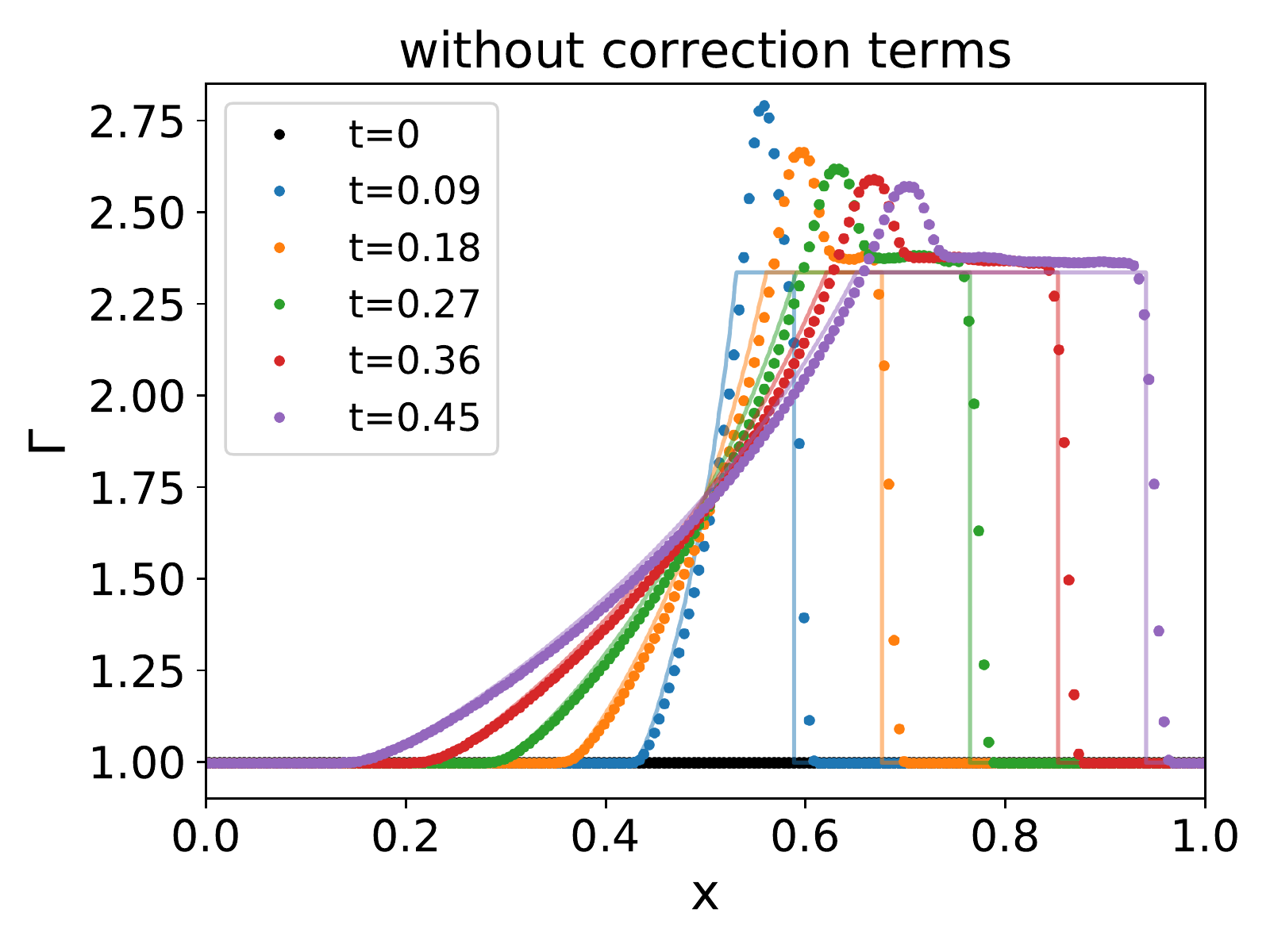}
\includegraphics[width=\columnwidth]{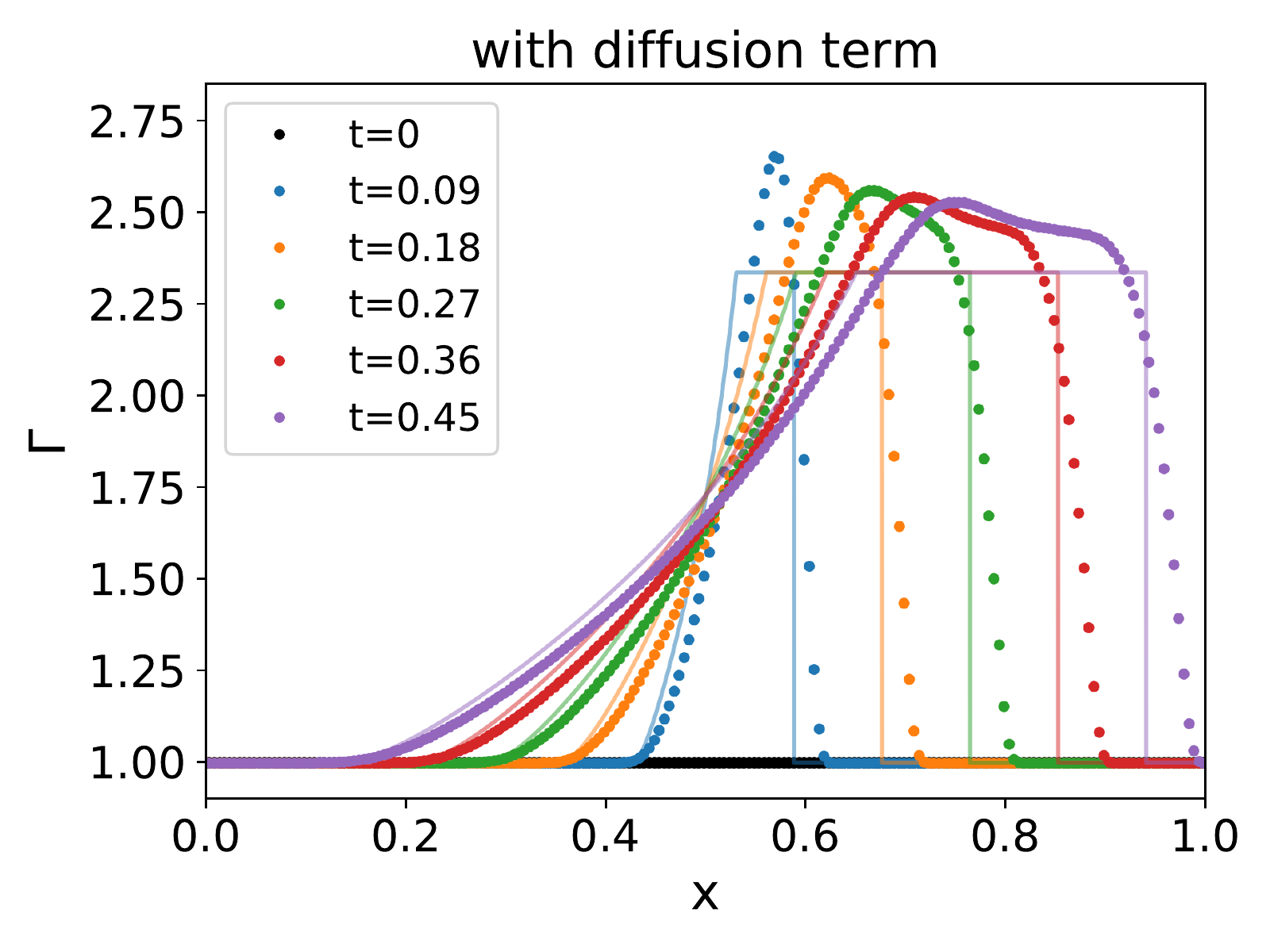}
\includegraphics[width=\columnwidth]{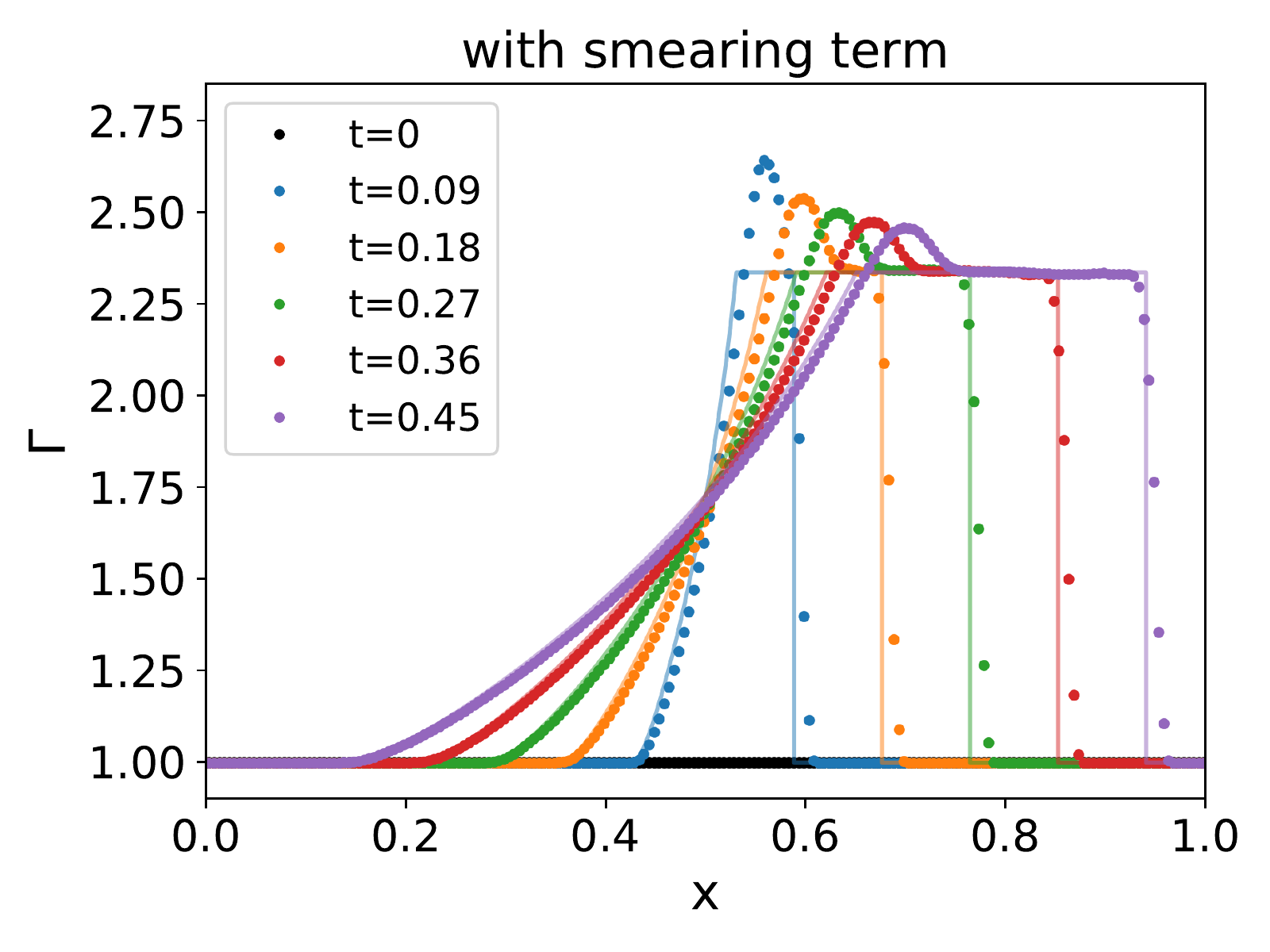}
\caption{
The upper panel illustrates the overshooting of the Lorentz factor at the upstream side of the high Lorentz factor plateau, i.e. before the rarefaction phase. The two lower ones are the same but with correction terms, i.e. diffusion and `smearing', as described in section~\ref{sec:overshooting}. The time-implicit scheme with a fixed equally spaced grid of $N=200$ was used. The initial conditions are $P_\mathrm{left} = 1000$, $\rho_\mathrm{left}=1$, $P_\mathrm{left} /P_\mathrm{right} = 5 \cdot 10^2$ and $\rho_\mathrm{left}/\rho_\mathrm{right} = 1$. The panel in the middle demonstrates the effect of the diffusion term. The corresponding parameters are chosen as $D_\mathrm{diff, D} = 0.0$, $D_\mathrm{diff, M_x} = 0.3$ and $D_\mathrm{diff, {\cal E}^d} = 0.3$. The lower panel demonstrates the effect of the `smearing' term using $\zeta = 0.1$.\label{fig:LF_overshooting}}
\end{figure}

\begin{figure*}
\centering

\tikzstyle{startstop} = [rectangle, rounded corners, minimum width=3cm, minimum height=1cm,text centered, draw=black, fill=red!30]
\tikzstyle{io} = [trapezium, trapezium left angle=70, trapezium right angle=110,minimum width=3cm, minimum height=1cm, text centered, text width =8em, draw=black, fill=blue!30]
\tikzstyle{process} = [rectangle, minimum width=3cm, minimum height=1cm, text centered, text width=8em, draw=black, fill=orange!30]
\tikzstyle{decision} = [diamond, minimum width=2.5cm, text centered, aspect=2, text width=4.1em, draw=black, fill=green!30]
\tikzstyle{arrow} = [thick,->,>=stealth]
\tikzstyle{darrow} = [thick,<->,>=stealth]

\begin{tikzpicture}[node distance=1.3cm]
\draw[dashed] (-2.0,-2.2) rectangle (2.0,-6.6);
\draw[dashed] (3.3,1.0) -- (14.9,1.0) -- (14.9,-7.5) -- (10.75,-7.5) -- (10.75,-9.5) -- (3.3,-9.5) -- cycle;
\node (start) [startstop] {Start};
\node (read)  [io, below of=start, yshift=0cm] {Read para\\-meter file}; 
\node (proId) [process, below of=read, yshift=-0.5cm] {Domain decomposition};
\node (proI) [process, below of=proId] {Initialize variables};
\node (proIr) [process, below of=proI] {Refine grid};
\node (proE0) [process, right of=start, xshift=4cm, yshift=0cm] {$t = t + \Delta t$};
\node (proE1) [process, below of=proE0] {Compute defect};
\node (proE2) [process, below of=proE1] {Construct preconditioner};
\node (proE3) [process, below of=proE2] {Invert matrices};
\node (proE4) [process, below of=proE3] {Compute pri\-mi\-tive variables};
\node (decC) [decision, below of=proE4, yshift=-0.25cm] {Converged?};
\node (deco) [decision, below of=decC, yshift=-0.5cm] {Time for output?};
\node (decf) [decision, right of=deco, xshift=2.5cm] {$t \geq t_\mathrm{final}$?};
\node (decR) [decision, right of=proE0, xshift=2.5cm] {Refine grid?};

\node (proR0) [process, right of=decR, xshift = 2.5cm, yshift=0.0cm] {Refine grid};
\node (proR1) [process, below of=proR0] {Compute main variables};
\node (proR2) [process, below of=proR1] {Update primitive variables};
\node (proR3) [process, below of=proR2] {Set global timestep $\Delta t$};
\node (proRd) [process, below of=proR3] {Update domain decomposition};
\node (write) [io, left of=deco, xshift=-4cm, yshift=0cm] {Write\\output};
\node (stop) [startstop, right of=decf, xshift=2.5cm] {Stop};

\draw [arrow] (start) -- (read);
\draw [arrow] (read) -- (proId);
\draw [arrow] (proId) -- (proI);
\draw [darrow] (proI) -- (proIr);
\draw [arrow] (proI) -| ++(2.65cm,2cm) |- (proE0);
\draw [arrow] (proI) -| ++(2.65cm,-2cm) |- ++(-1cm,-0.9cm) -| (write);
\draw [arrow] (proE0) -- (proE1);
\draw [arrow] (proE1) -- (proE2);
\draw [arrow] (proE2) -- (proE3);
\draw [arrow] (proE3) -- (proE4);
\draw [arrow] (proE4) -- (decC);
\draw [arrow] (decC) -| node[anchor=north] {no} ++(2.5cm,2cm) |- (proE1);

\draw [arrow] (decC) -- node[anchor=east] {yes}(deco);
\draw [darrow] (deco) -- node[anchor=south] {yes} (write);
\draw [arrow] (deco) -- node[anchor=south] {no} (decf);
\draw [arrow] (decf) -- node[anchor=south] {yes} (stop);
\draw [arrow] (decf) -- node[anchor=east, yshift=-3.1cm] {no} (decR);
\draw [arrow] (decR) -- node[anchor=south] {no} (proE0);
\draw [arrow] (decR) -- node[anchor=south] {yes} (proR0);

\draw [arrow] (proR0) -- (proR1);
\draw [arrow] (proR1) -- (proR2);
\draw [arrow] (proR2) -- (proR3);
\draw [arrow] (proR3) -- (proRd);
\draw [arrow] (proRd) |- ++(-3.8cm,-1.5cm);
\end{tikzpicture}
\caption{
A flow chart snapshot of the algorithm of the time-implicit solution 
procedure with AMR. Starting with reading the parameter file and building the initial conditions on a refined grid. The steps therefore are displayed in the left dashed box. The right dashed box shows the core routines of the code, which evolve the solution in time and adapt the grid after a couple of time steps.
Outputs are written according to the specified times in the parameter file and also the initial setup is outputted. The computer code is written in C++. \label{fig:flow_chart}}
\end{figure*}
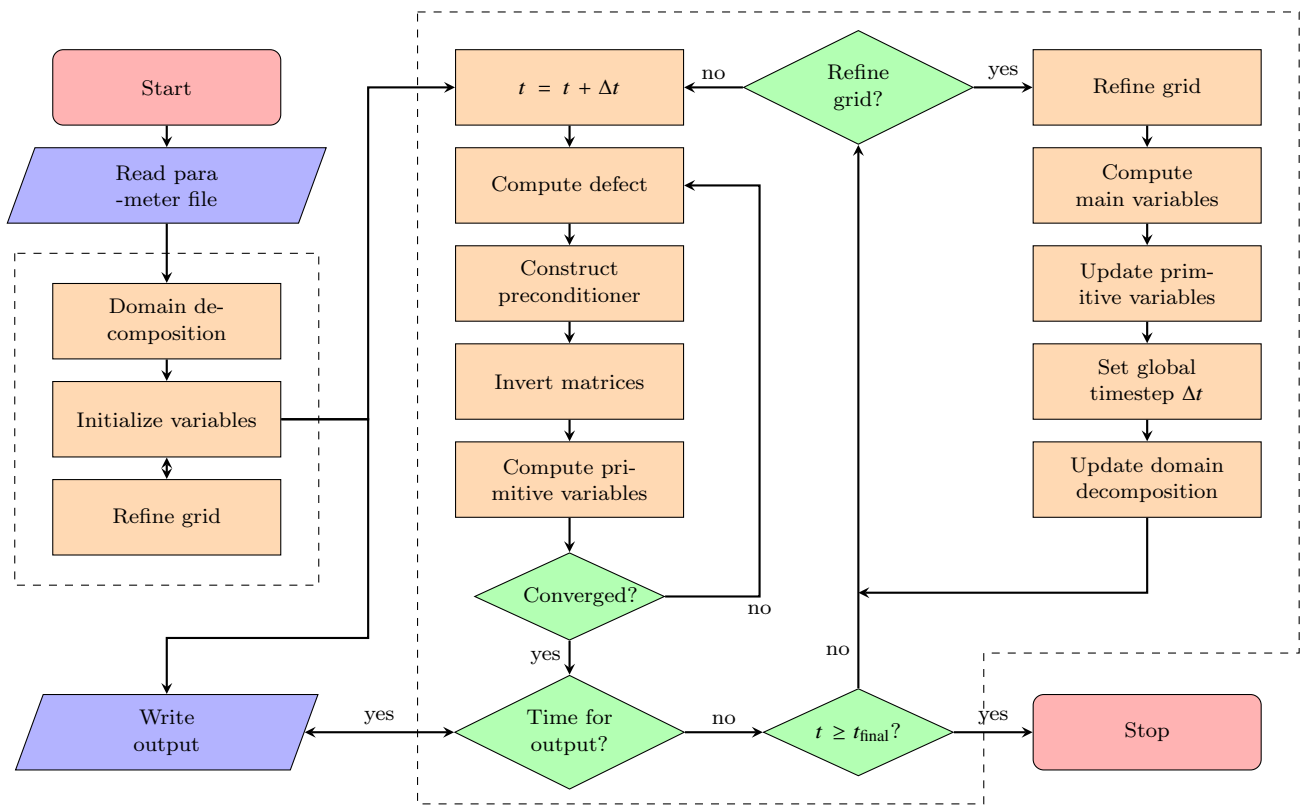

\subsection{How to prevent overshooting?} \label{sec:overshooting}

We use different techniques to reduce the overshooting (in Lorentz factor, not the classical well-known problem in terms of density right after the shock front), which occurs at the upstream side of the high Lorentz factor plateau (see upper panel of Fig.~\ref{fig:LF_overshooting}). This problem is hardest when the shock front just arises, the thinner the shock front the larger the overshooting. Unfortunately, this problem can't be reduced by using a higher resolution. There will be still a few cells, which face this problem. This is because the density pile-up starts as an infinitely small one and grows larger in time. How can we treat the very first time of the simulation when the shock front arises? As changing the resolution does not lead to a major improvement we alter the formulation to make the code more capable of this situation. We have two approaches to this problem. One is based on adding a diffusion term and the other one is based on `smearing out' the formulation.

\subsubsection{Diffusion}
In general diffusion of $q$ is given by:
\begin{equation}
Q_\mathrm{diff} = \nabla \cdot (D_\mathrm{diff} \, \nabla q) \, .
\end{equation}
The implementation contains such a term for all three relativistic Euler equations.
A simple ansatz is to use a constant diffusion coefficient, as we did in our simulations. However, one could build a model with a variable coefficient that tries to capture numerical difficult regions.

The middle panel of Fig.~\ref{fig:LF_overshooting} demonstrates the effect of the constant diffusion term. Compared to the upper panel without correction terms, it reduces the overshooting but also makes the shock front less sharp. Consequently, for our simulations, we prefer to choose the diffusion parameter as small as possible.

\subsubsection{`Smearing'}
We alter the derivative $\partial (u^t V^x)^n_j / \partial x$ in the energy equation. 
This method is based on the idea to spatially `smear out' a bit the $\partial (u^t V^x)^n_j / \partial x$ derivative. The idea is to introduce a dependence on neighbouring cells. This can be done as follows:
\begin{flalign} \label{eq:smearing}
 \frac{\Delta (u^t V^x)_{j}}{\Delta x_j} =& \, u^t_j \frac{\Delta V^x_j}{\Delta x_j} + \frac{V^x_j}{\Delta x_j} \, \bigg \{ (1 - \zeta) \, \Delta u^t_j &&\nonumber\\
 &+ \frac{\zeta}{2} \left[ (1 + \Theta_j) \, \Delta u^t_{j+1/2} + (1-\Theta_j) \Delta u^t_{j-1/2} \right] \bigg\} \, . &&
\end{flalign}
We only add $u^t$ from the downstream side. Note, therefore $\Theta_j = \Theta(V^x_j)$ is used. The `smearing' factor $\zeta$ controls how much the solution depends on the downstream neighbour value of $u^t$. If $\zeta$ equals zero it is independent and if $\zeta$ equals one the solution relies only on the downstream value. Note, this also suppresses the classical over- and undershooting of the density, that occurs right after the shock front.

In the lower panel of Fig.~\ref{fig:LF_overshooting}, we demonstrate the effect of the `smearing' term. Compared to the upper panel without correction terms, it reduces the overshooting and lowers the high Lorentz factor plateau. Moreover, it does not reduce the sharpness of the shock front like the diffusion term.

\subsection{The algorithm}

In Fig.~\ref{fig:flow_chart} we map the flow chart of our code for the time-implicit scheme with AMR. The code is written in C++ and parallelised for shared memory architectures. Here, we discuss the technicalities of the AMR implementation and the parallelisation.

\subsubsection{Adaptive mesh}
We make use of the h-refinement strategy, which is applied in many other codes. The applied strategy has the advantage that not all cells need to be modified when the grid is adapted to the physical problem. Only the variables of cells involved in refinement or derefinement change.
This is in contrast to r-refinement, where the number of grid points stays constant, but they are relocated to follow the interesting features of the physical solution.
Furthermore, the use of discrete refinement levels is advantageous when calculating the derivatives. This is because the derivatives depend on the ratios of the cell sizes to each other. Mainly a couple of different ratios of cell sizes are involved instead of recomputing every time how much a cell contributes to a derivative one can speed up the code by hard-coding them.
For the implementation of the grid, we use an array, which is quite fine to do AMR in one dimension. Higher-dimensional codes \citep[see e.g.][]{Khokhlov_1998, Teyssier_2002, Fromang_2006} usually use other structures like the `Fully Threaded Tree' \citep{Khokhlov_1998} to store the data, as AMR is more complicated in higher dimensions.

\subsubsection{Parallelisation}
To parallelise the time-explicit scheme we solely make use of domain decomposition. This allows for a freely chosen number and size of domains. Where the first corresponds to the number of threads created. Thus domain decomposition allows for efficient parallelisation. In contrast, the parallelisation of the implicit scheme is more difficult. We chose to solve the relativistic Euler equations in parallel.
This approach limits the improvement of the parallelisation drastically as the number of threads, which can be processed in parallel is limited by the number of equations, in our case three.
However, the computation of the primitive variables is again parallelised using domain decomposition.
The parallelisation of our C++ code is realised for shared memory architectures by using pthreads (POSIX threads).
\section{Application: propagation of relativistic Shocks} \label{sec:results}
Sod's shock tube problem \cite[see][]{Sod_1978} has become the standard test problem for modelling the propagation of shocks in the Newtonian regime, where the velocities of propagation are far below the speed of light.
However, as $v \rightarrow c,$ relativistic effects become important and the transport operator should then be modified and corrected accordingly. Detailed information about the relativistic version of the shock tube problem can be found in \cite{Marti_2003}.
However, the obtained numerical results here are compared to the exact solution obtained according to \cite{Thompson_1986}.

We consider a variety of relativistic shock tube problems with different initial conditions. All problems we consider have in common that the initial transport velocity $V^x=0$ and that the left state is given by $P_L=1000$ and $\rho_L=1$.
Furthermore, for closing the system of equations, the equation of state for a perfect gas is assumed with an adiabatic index of $\gamma = 5/3$.

To enhance efficiency and spatial accuracy of our numerical algorithm, we employ an adaptive mesh refinement strategy. Here the gradients of the relativistic density $D$ and the pressure $P$ are used as (de)refinement criteria.

\subsection{Spectrum of Lorentz factors} \label{sec:spec_LF}

To compare the different schemes qualitatively we did a couple of simulations for two different initial conditions. First, with $P_\mathrm{left}/P_\mathrm{right}=10^{2}$ and $\rho_\mathrm{left}/\rho_\mathrm{right}=1$, which lead to a Lorentz factor of $\Gamma = 1.7$. The second initial conditions are $P_\mathrm{left}/P_\mathrm{right}=10^{5}$, $\rho_\mathrm{left}/\rho_\mathrm{right}=1$, which lead to a Lorentz factor of $\Gamma = 3.59$. Note, these are the same initial conditions as considered for the first time by \cite{Norman_1986} to study high Lorentz factors. Besides, we study even more extreme initial conditions. The corresponding results are presented in sections~\ref{sec:very_high_LF} and \ref{sec:con_eff}.


\begin{figure}
\includegraphics[width=\columnwidth]{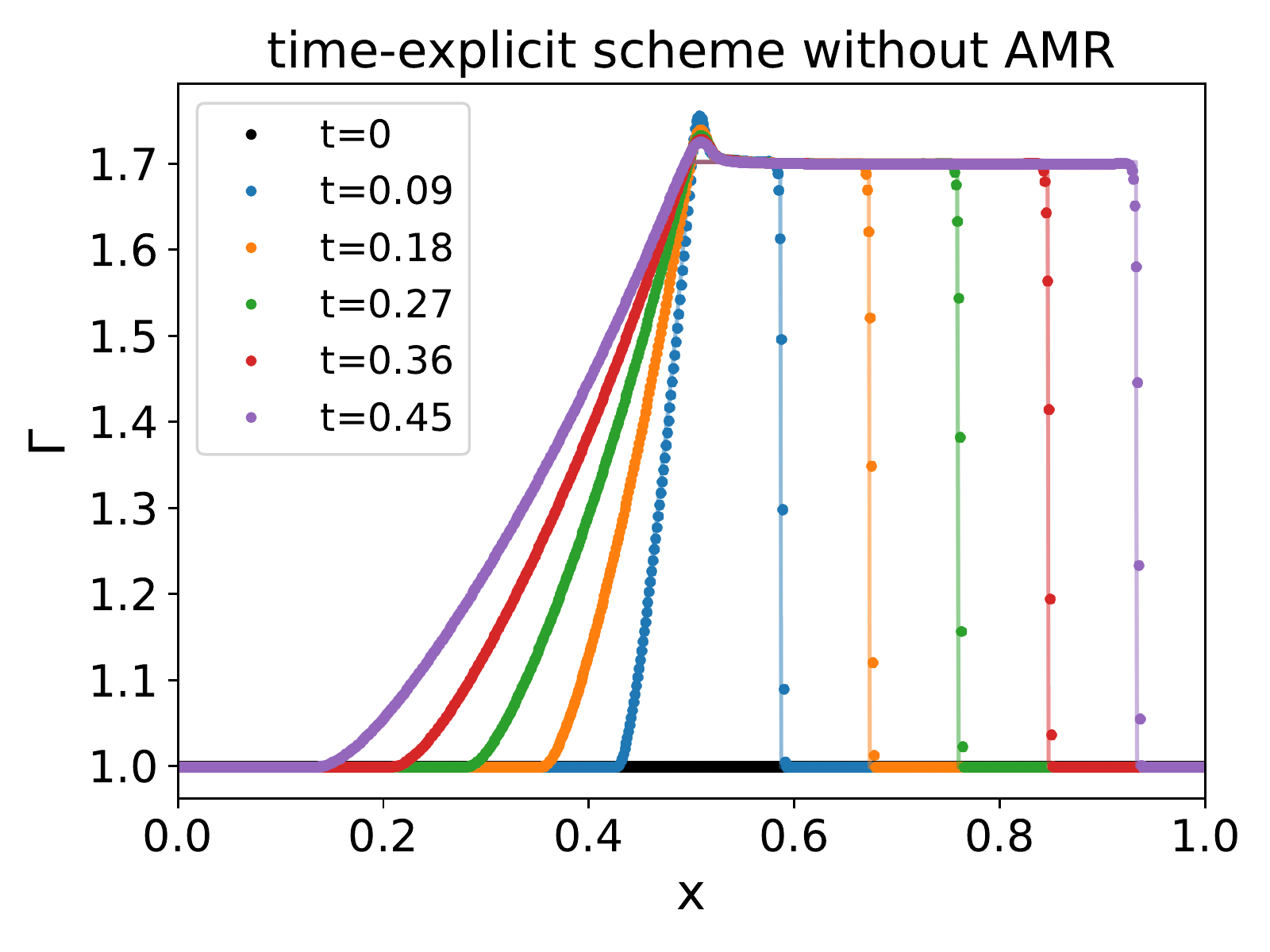}
\includegraphics[width=\columnwidth]{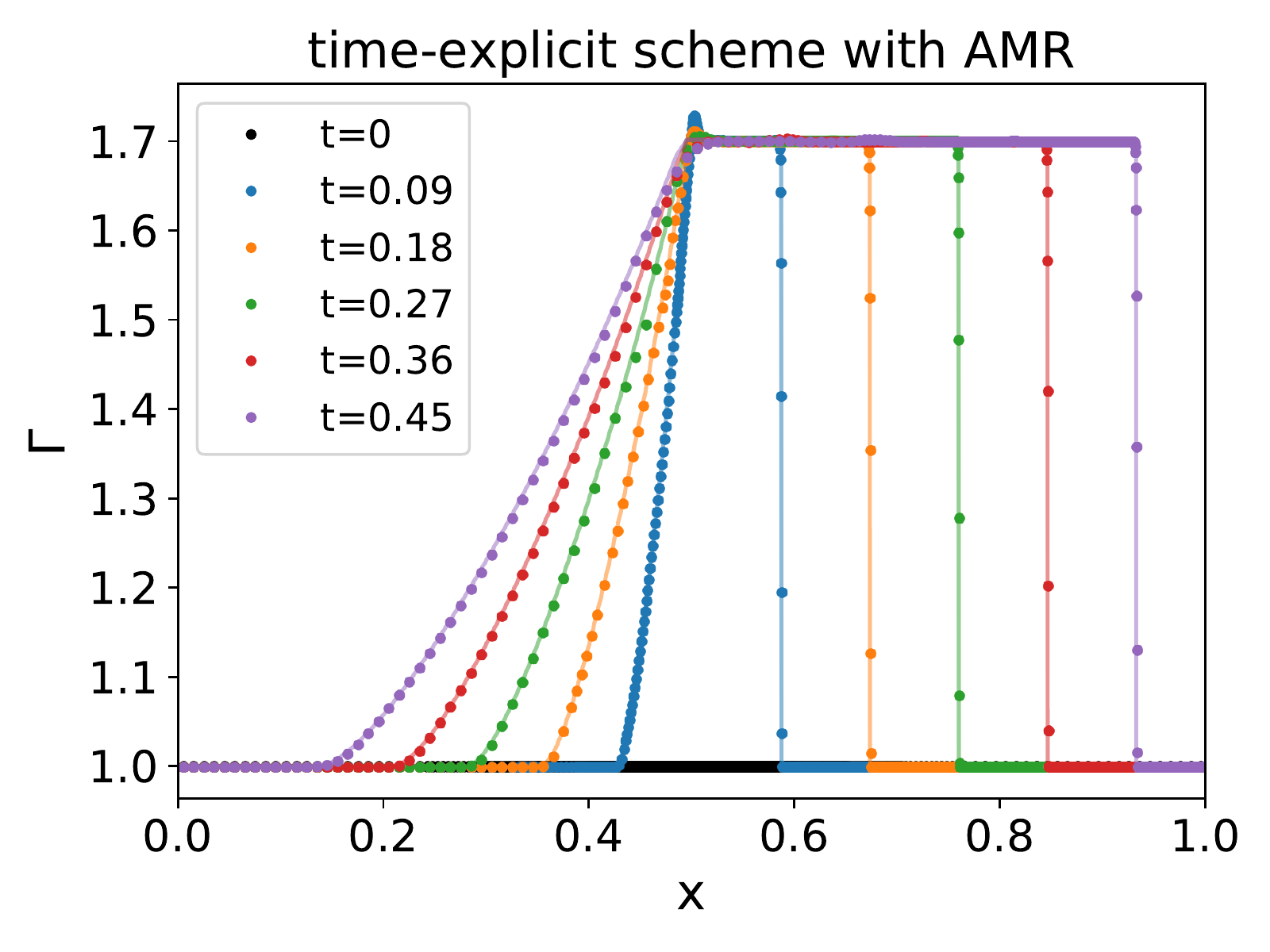}
\caption{The spatial distribution of the Lorentz factor (dots) compared to the exact solution (lines) using the time-explicit scheme with and without AMR for different elapsed times is shown.
These calculations are based on $P_\mathrm{left}/P_\mathrm{right}=10^{2}$ and $\rho_\mathrm{left}/\rho_\mathrm{right}=1$ initially.
For the run without AMR, $N_\mathrm{no~AMR} = 800$ cells have been used and $N_\mathrm{AMR} \sim 740$ cells with AMR. These runs have been performed without artificial viscosity, but $\zeta = 0.2$ was chosen. 
AMR diffusion parameters are chosen as $D_{\mathrm{diff,}D}=0.0$, $D_{\mathrm{diff,}M_x}= 1.5625 \cdot 10^{-6}$, $D_{\mathrm{diff,}{\cal E}^d} = 1.5625 \cdot 10^{-6}$ and without AMR it is $D_{\mathrm{diff,}D}=0.0$, $D_{\mathrm{diff,}M_x}= 6.25 \cdot 10^{-6}$, $D_{\mathrm{diff,}{\cal E}^d} = 6.25 \cdot 10^{-6}$.
\label{fig:LF_r1_exp}}
\end{figure}

\begin{figure}
\includegraphics[width=\columnwidth]{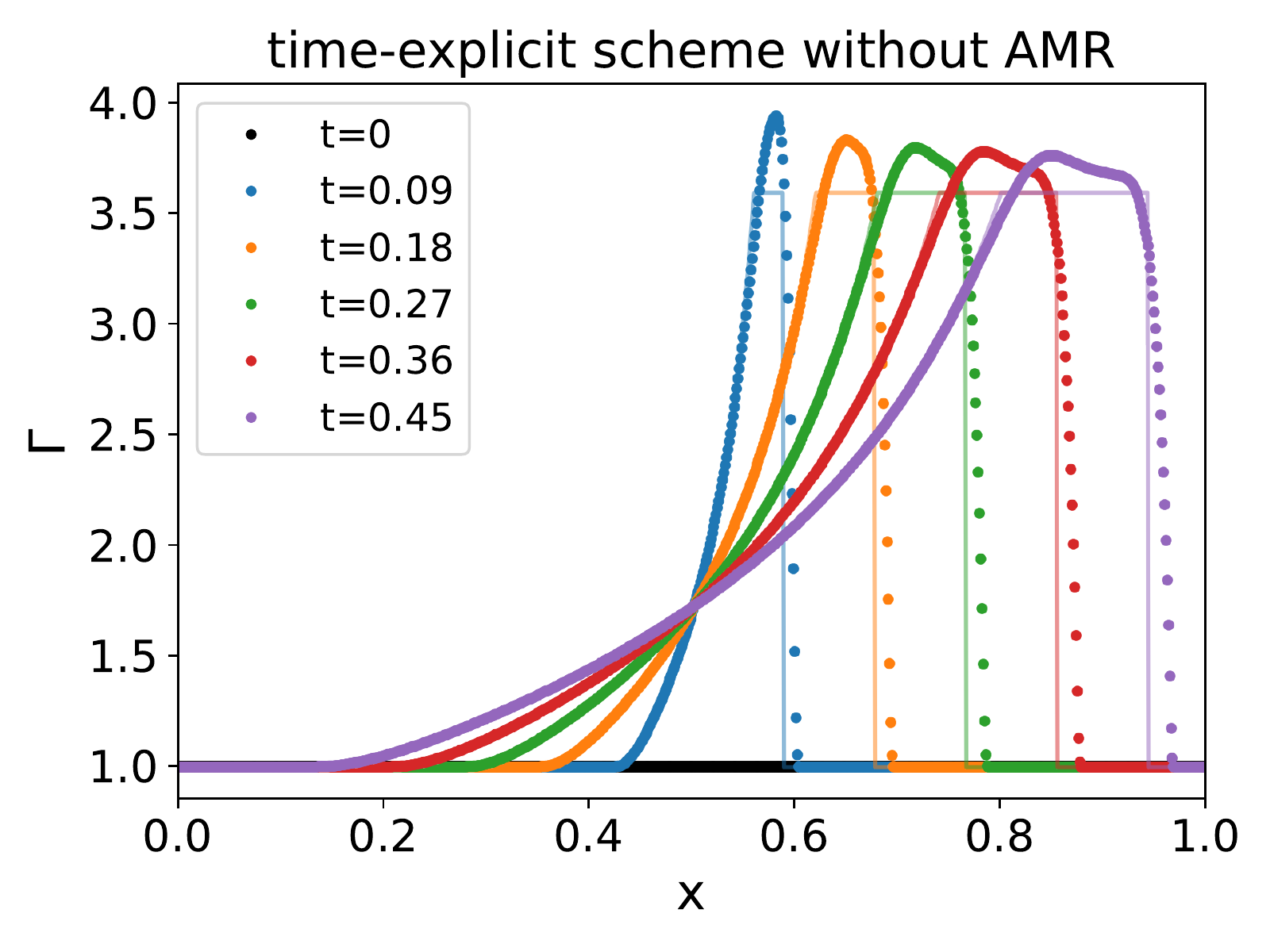}
\includegraphics[width=\columnwidth]{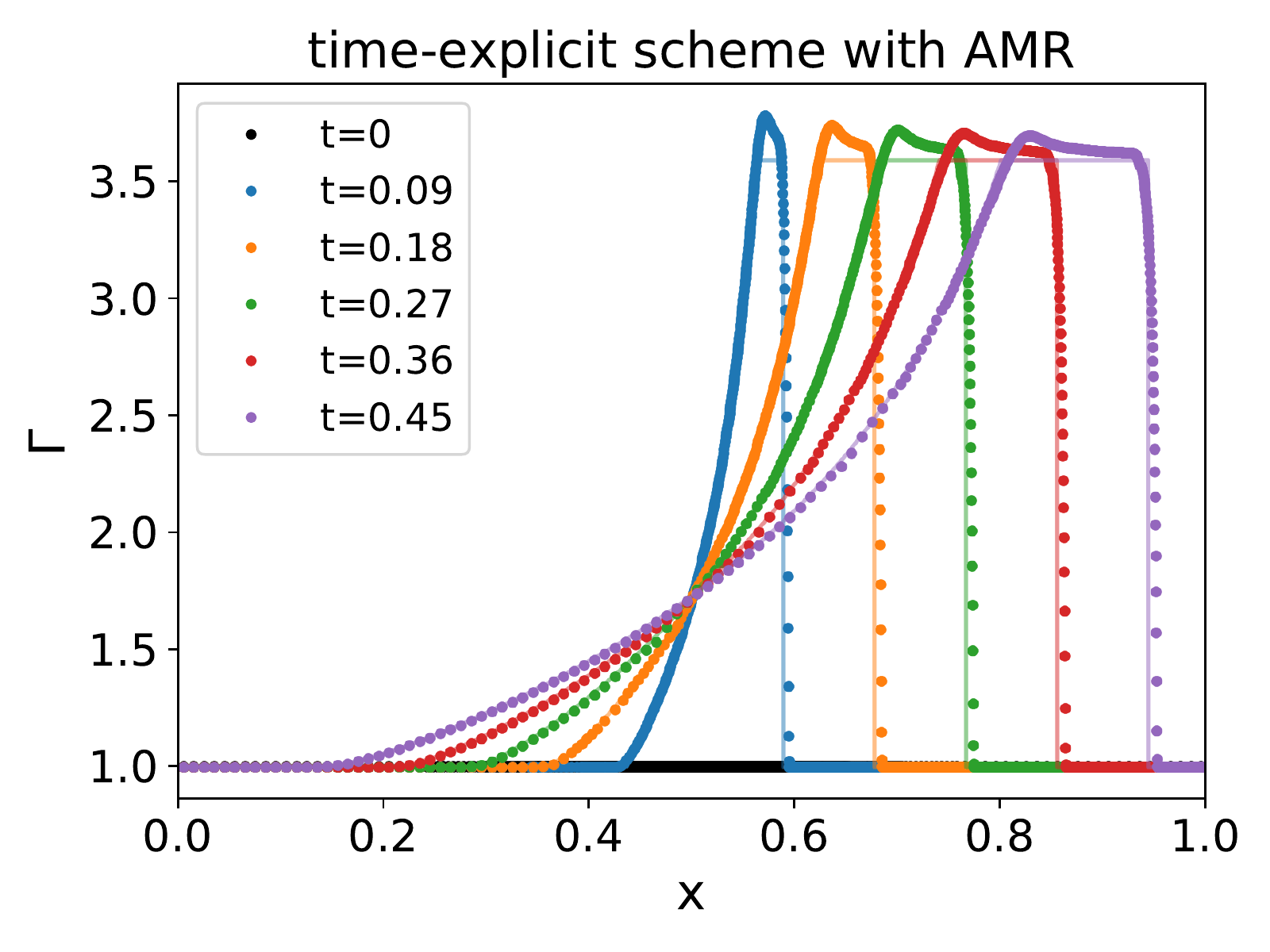}
\caption{The same as in the previous plot (Fig.~\ref{fig:LF_r1_exp}), though the runs are based on using the initial jump:
$P_\mathrm{left}/P_\mathrm{right}=10^{5}$ and $\rho_\mathrm{left}/\rho_\mathrm{right}=1$.
The numbers of grid cells used are: $N_\mathrm{no~AMR} = 800$ and $N_\mathrm{AMR} \sim 700$.
For the AMR run the diffusion parameters are chosen as $D_{\mathrm{diff,}D}=0.0$, $D_{\mathrm{diff,}M_x}= 5 \cdot 10^{-5}$, $D_{\mathrm{diff,}{\cal E}^d} = 5 \cdot 10^{-5}$ and without AMR it is $D_{\mathrm{diff,}D}=0.0$, $D_{\mathrm{diff,}M_x}= 2 \cdot 10^{-4}$, $D_{\mathrm{diff,}{\cal E}^d} = 2 \cdot 10^{-4}$.
Furthermore, the other parameters are chosen as follows: $\alpha_\mathrm{art}=6$ and $\zeta = 0.5$. \label{fig:LF_r2_exp}}
\end{figure}
\begin{figure}
\includegraphics[width=\columnwidth]{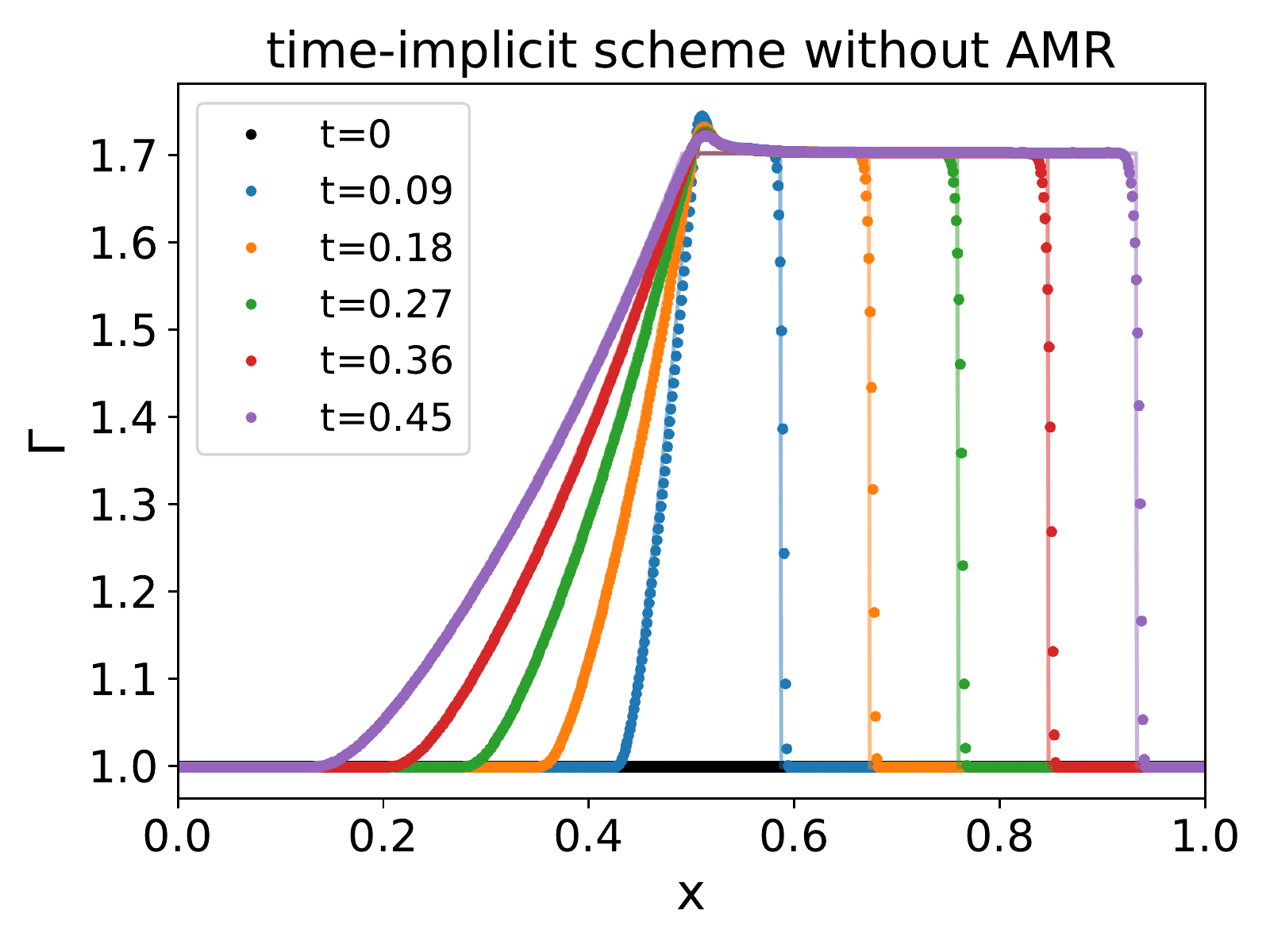}
\includegraphics[width=\columnwidth]{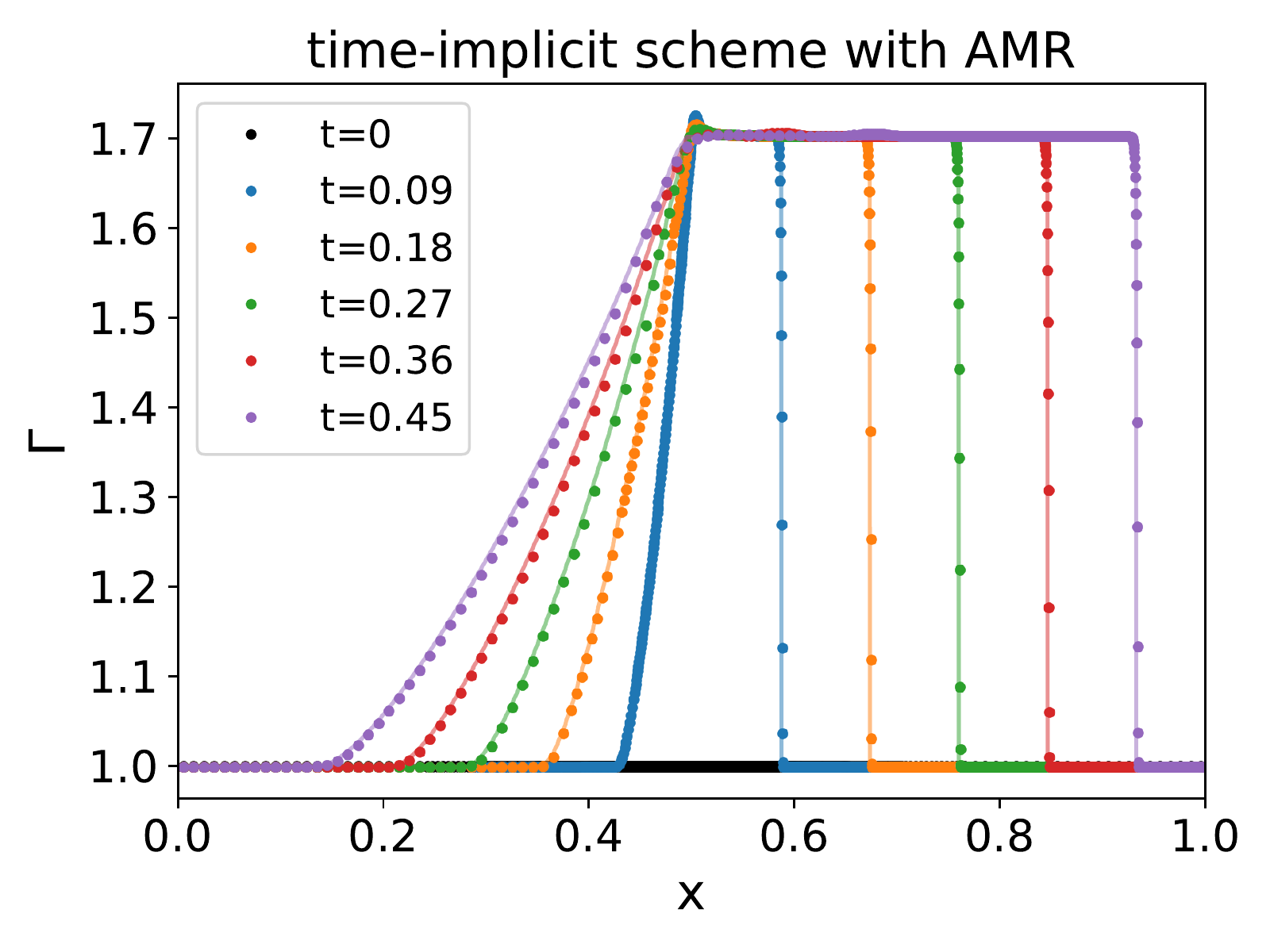}
\caption{
The same as in Fig.~\ref{fig:LF_r1_exp}, though the solution has been obtained using the time-implicit approach.
The following parameters were chosen to be non-zero: 
$D_{\mathrm{diff,}D}=0.0$, $D_{\mathrm{diff,}M_x}= 0.1$, $D_{\mathrm{diff,}{\cal E}^d} = 0.1$.
Although $\mathcal{C}_\mathrm{CFL}^\mathrm{implicit} $ here is much larger than ${C}_\mathrm{CFL}^\mathrm{explicit}$ in Fig. \ref{fig:LF_r1_exp},
the difference between the results of both methods appears to be negligibly small.
\label{fig:LF_r1_imp}
}
\end{figure}

\begin{figure}
\includegraphics[width=\columnwidth]{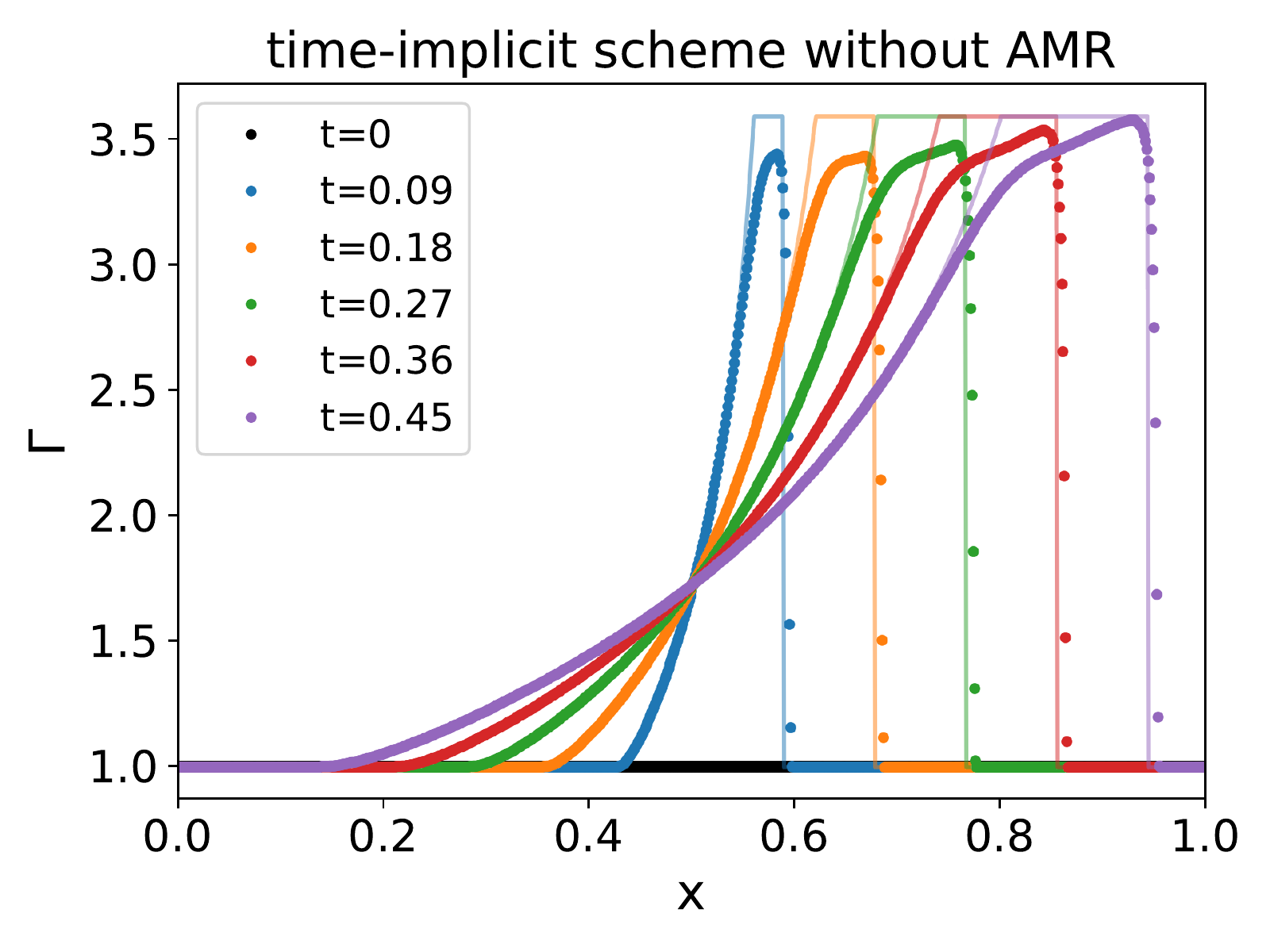}
\includegraphics[width=\columnwidth]{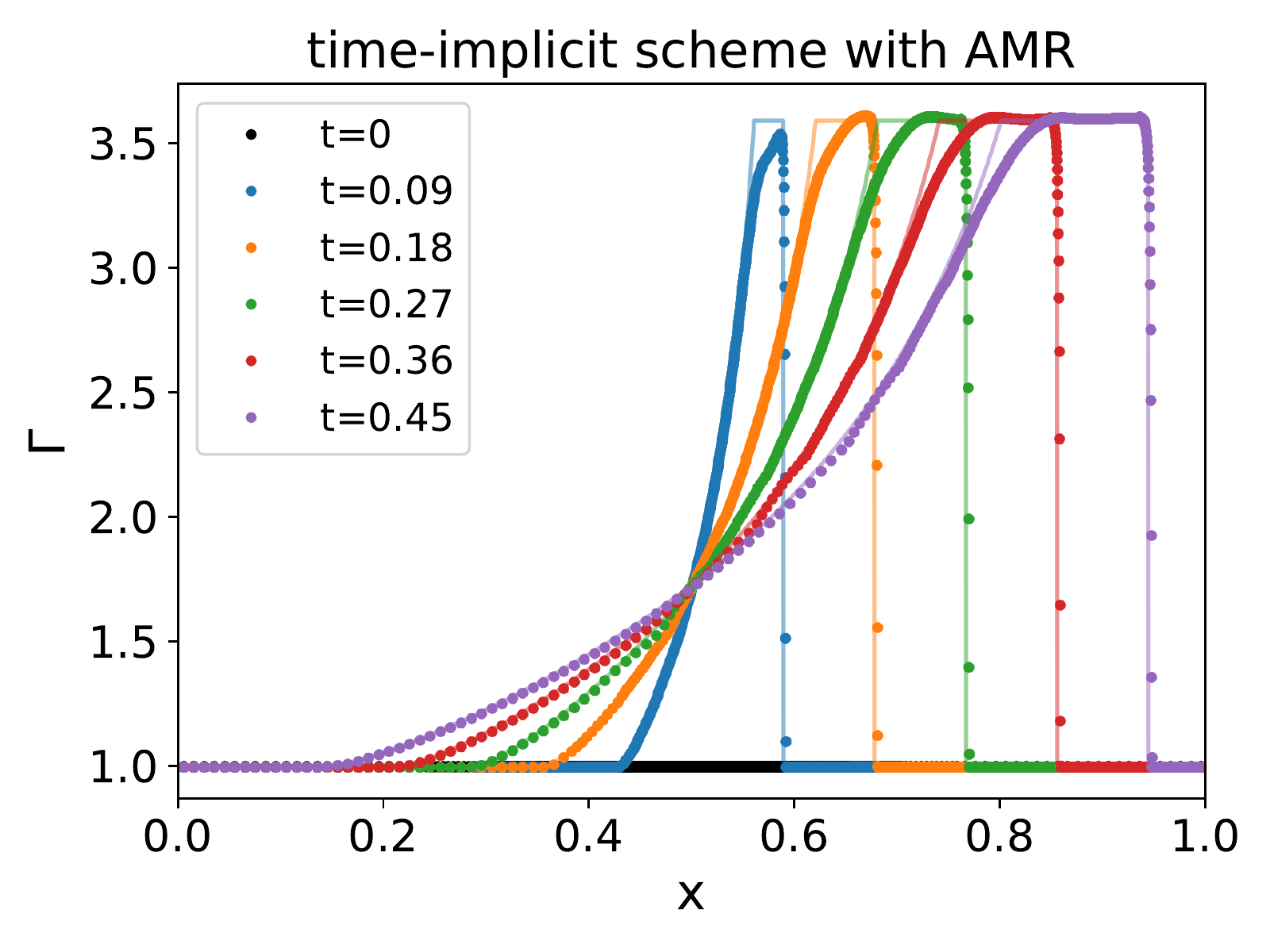}
\caption{
The same as in Fig.~\ref{fig:LF_r2_exp}, though the implicit scheme was used to evolve the initial conditions in time. Hence, different values for the following parameters are chosen: $\alpha_\mathrm{art}=0.0$, $D_{\mathrm{diff,}D}=0.0$, $D_{\mathrm{diff,}M_x}=0.07$, $D_{\mathrm{diff,}{\cal E}^d} = 0.07$ and $\zeta = 0.5$.
The number of grid cells, which has been used, are: $N_\mathrm{no~AMR} = 800$ and $N_\mathrm{AMR} \sim 700$. The differences between the results of both methods (compare to Fig.~\ref{fig:LF_r2_exp}) are larger than for the smaller Lorentz factor of $\Gamma=1.7$ (figures~\ref{fig:LF_r1_exp} and \ref{fig:LF_r1_imp}). Further quantities of the simulation shown in the lower panel are displayed in figures~\ref{fig:p1_i2_AMR} and \ref{fig:p2_i2_AMR}. \label{fig:LF_r2_imp}}
\end{figure}
\begin{figure}
\includegraphics[width=\columnwidth]{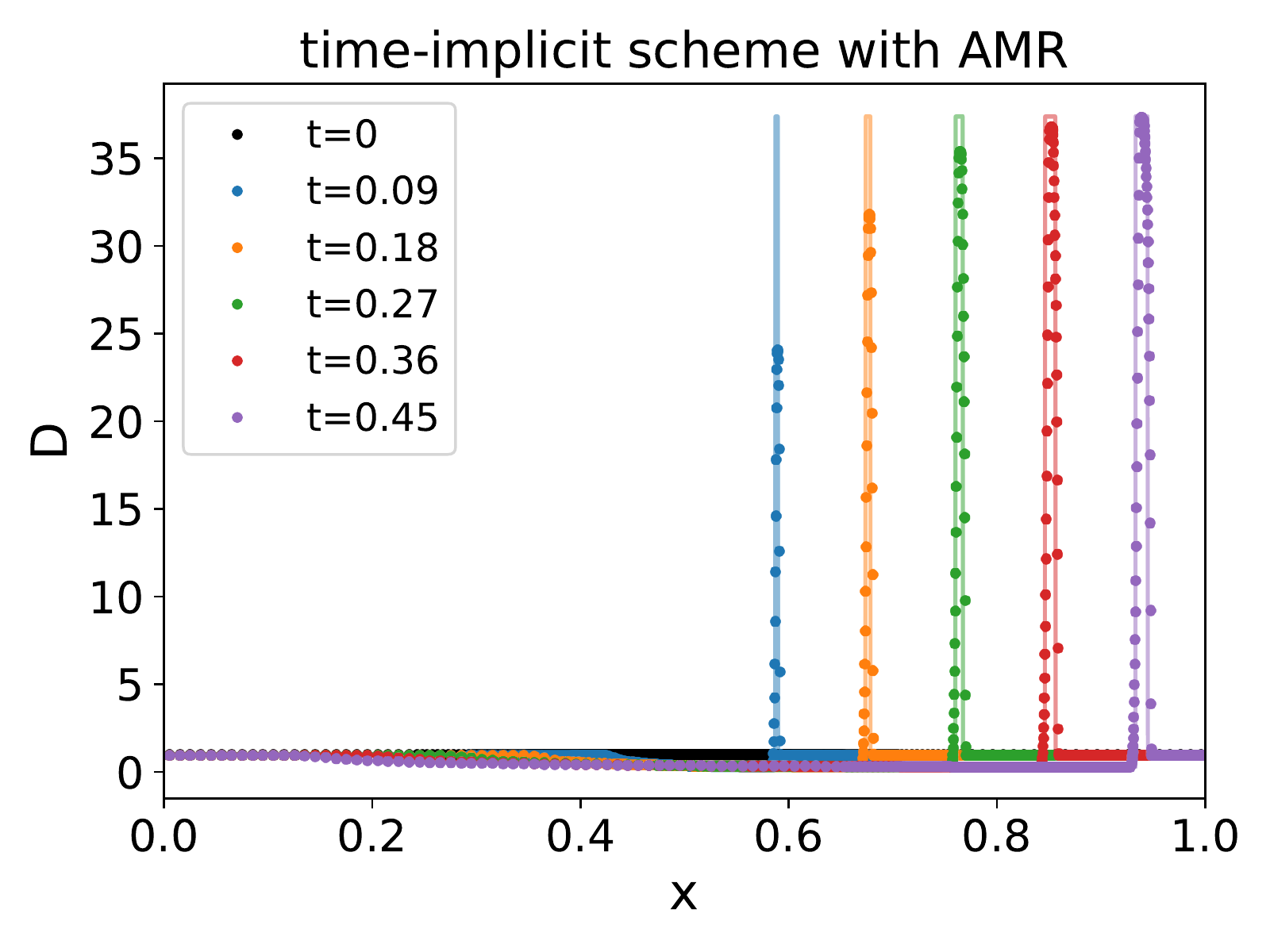}
\includegraphics[width=\columnwidth]{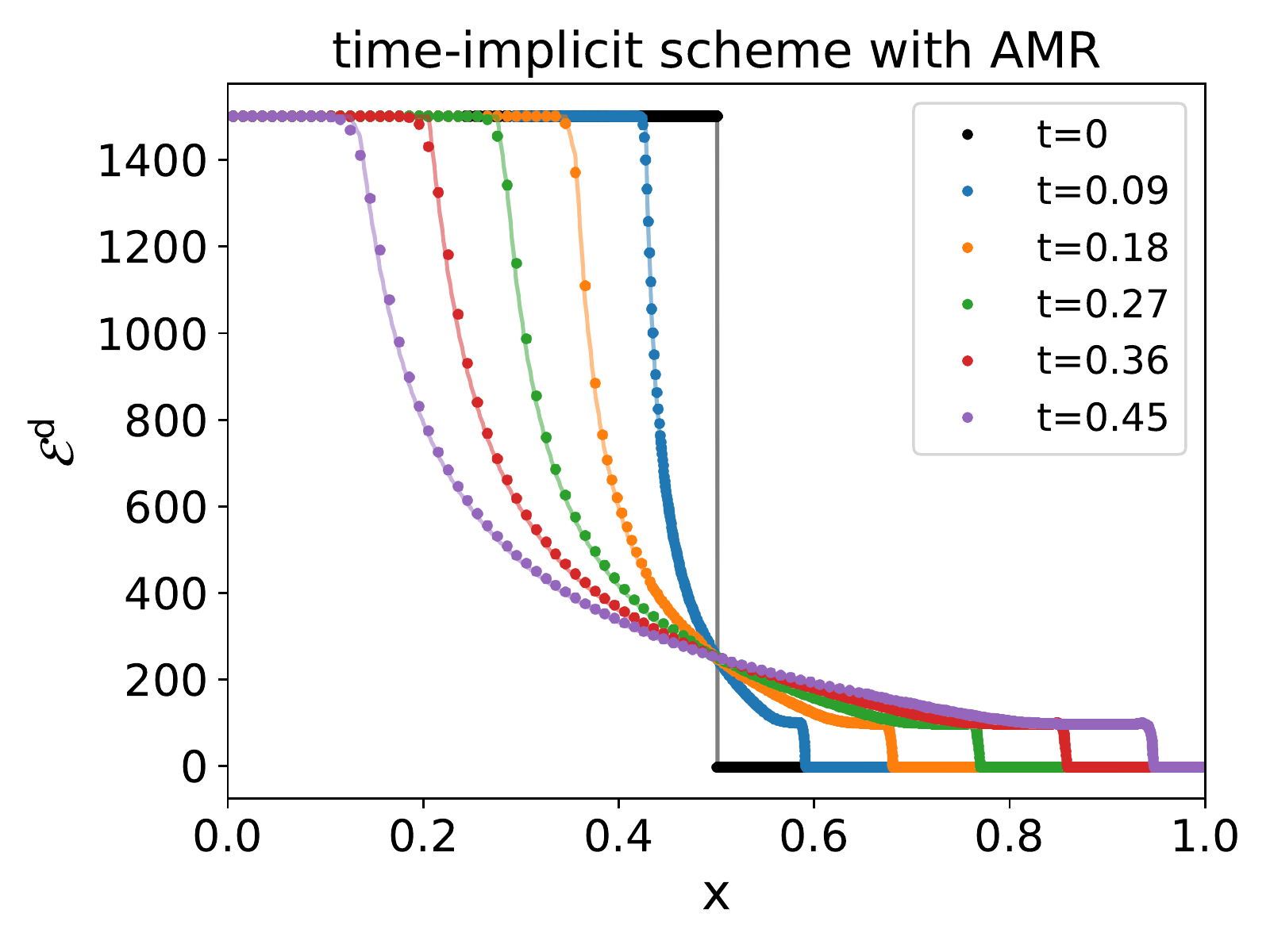}
\caption{The spatial distributions of the relativistic density $D$ (upper panel) and of ${\cal E}^d$ (lower panel) using the time-implicit solution procedure. Plotted for the same simulation as in the lower panel of Fig.~\ref{fig:LF_r2_imp}.
 \label{fig:p1_i2_AMR}}
\end{figure}

\begin{figure}
\includegraphics[width=\columnwidth]{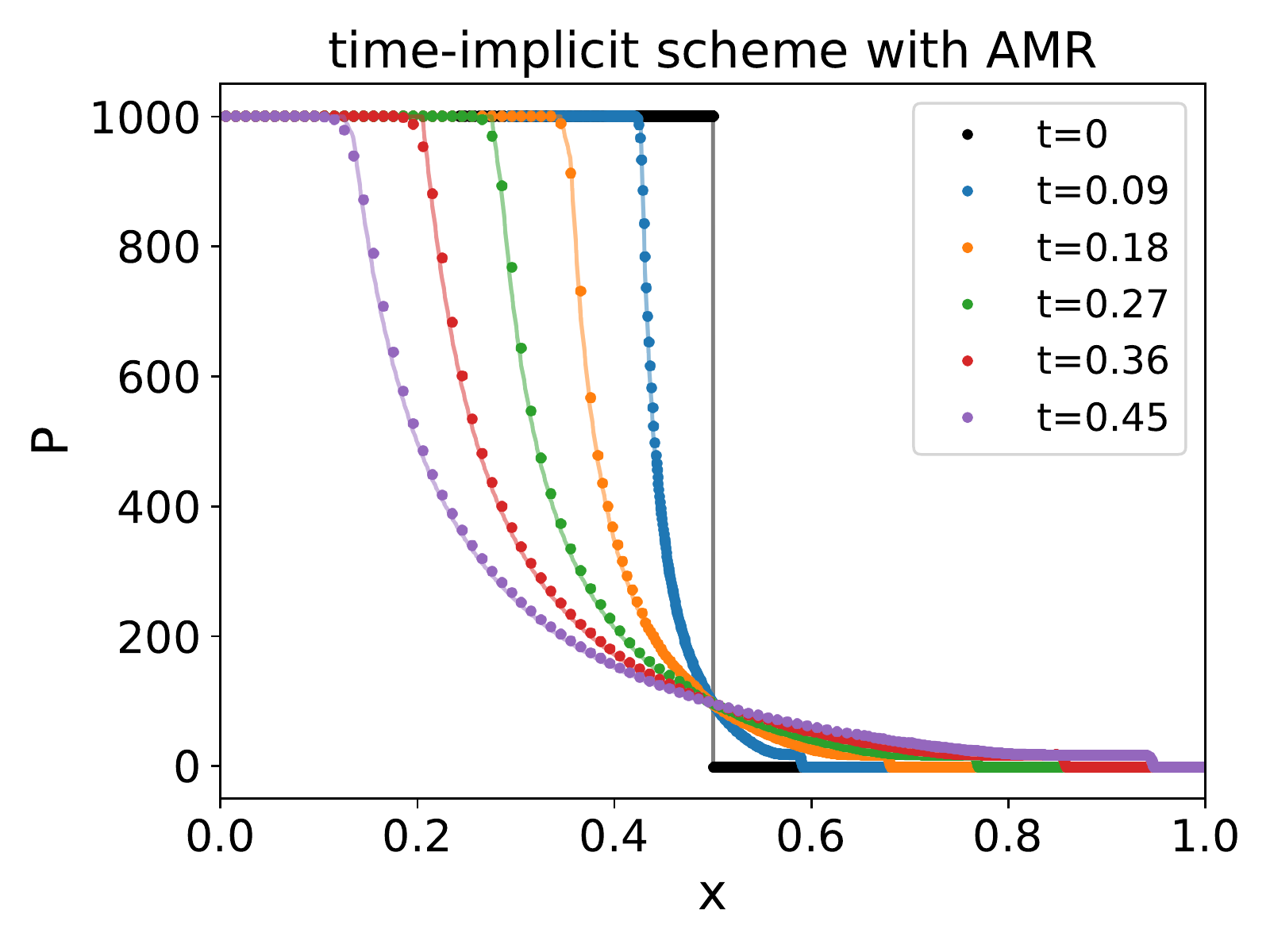}
\includegraphics[width=\columnwidth]{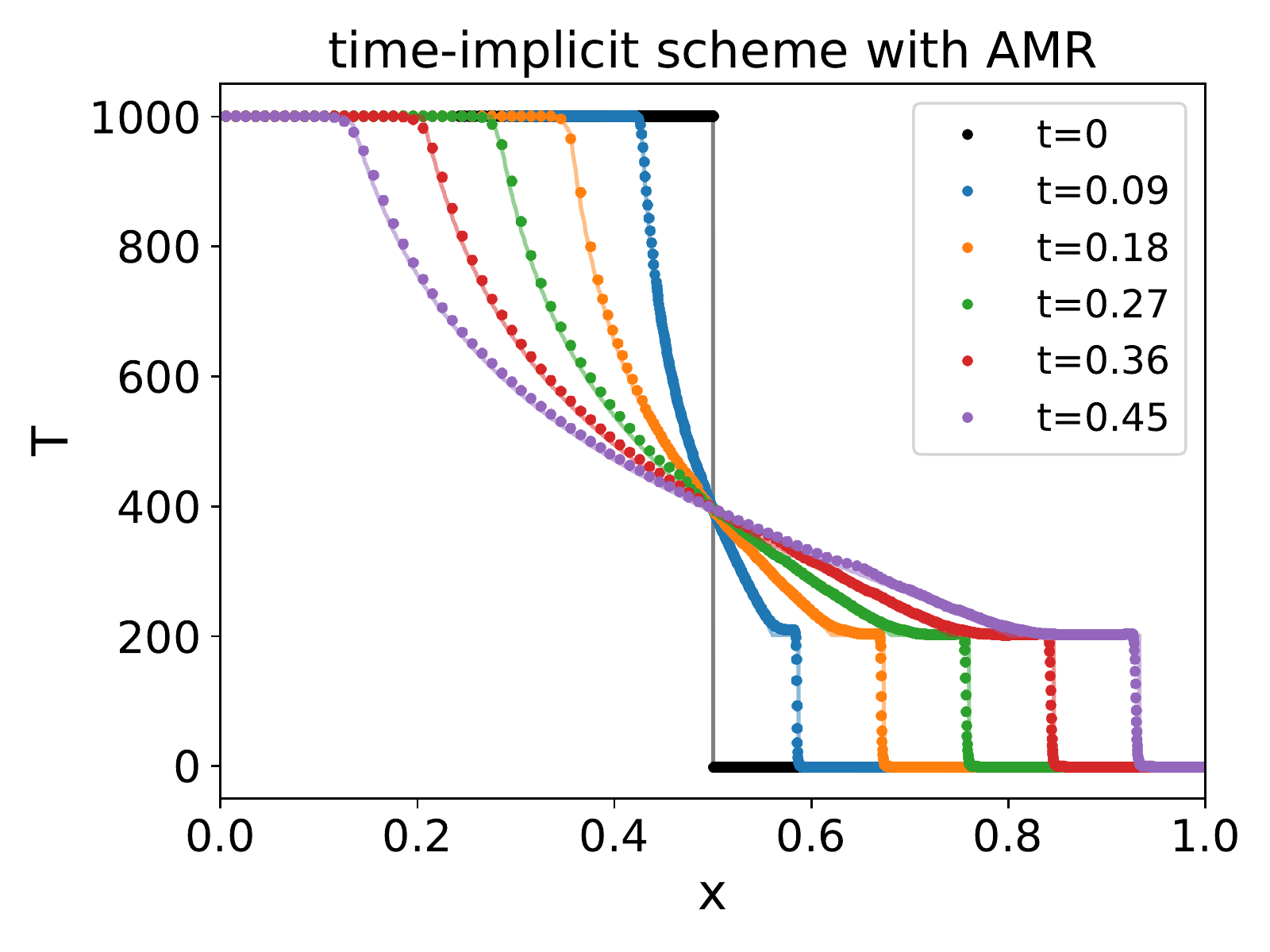}
\caption{The same as in the previous figure (Fig.~\ref{fig:p1_i2_AMR}), though now the spatial distributions of the pressure $P$ (upper panel) and the temperature $T$ are shown.\label{fig:p2_i2_AMR}
}
\end{figure}
\begin{figure}
\includegraphics[width=\columnwidth]{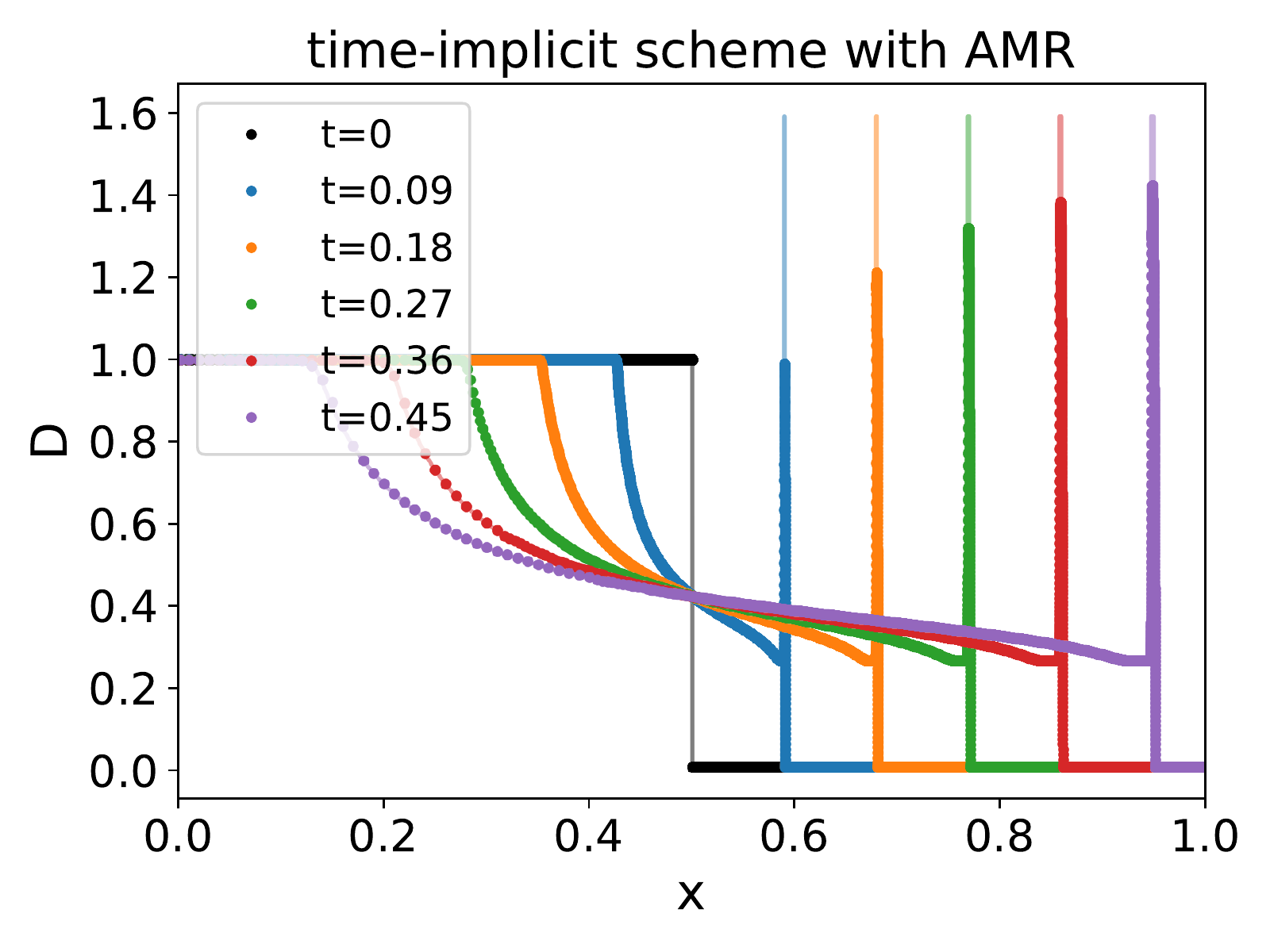}
\includegraphics[width=\columnwidth]{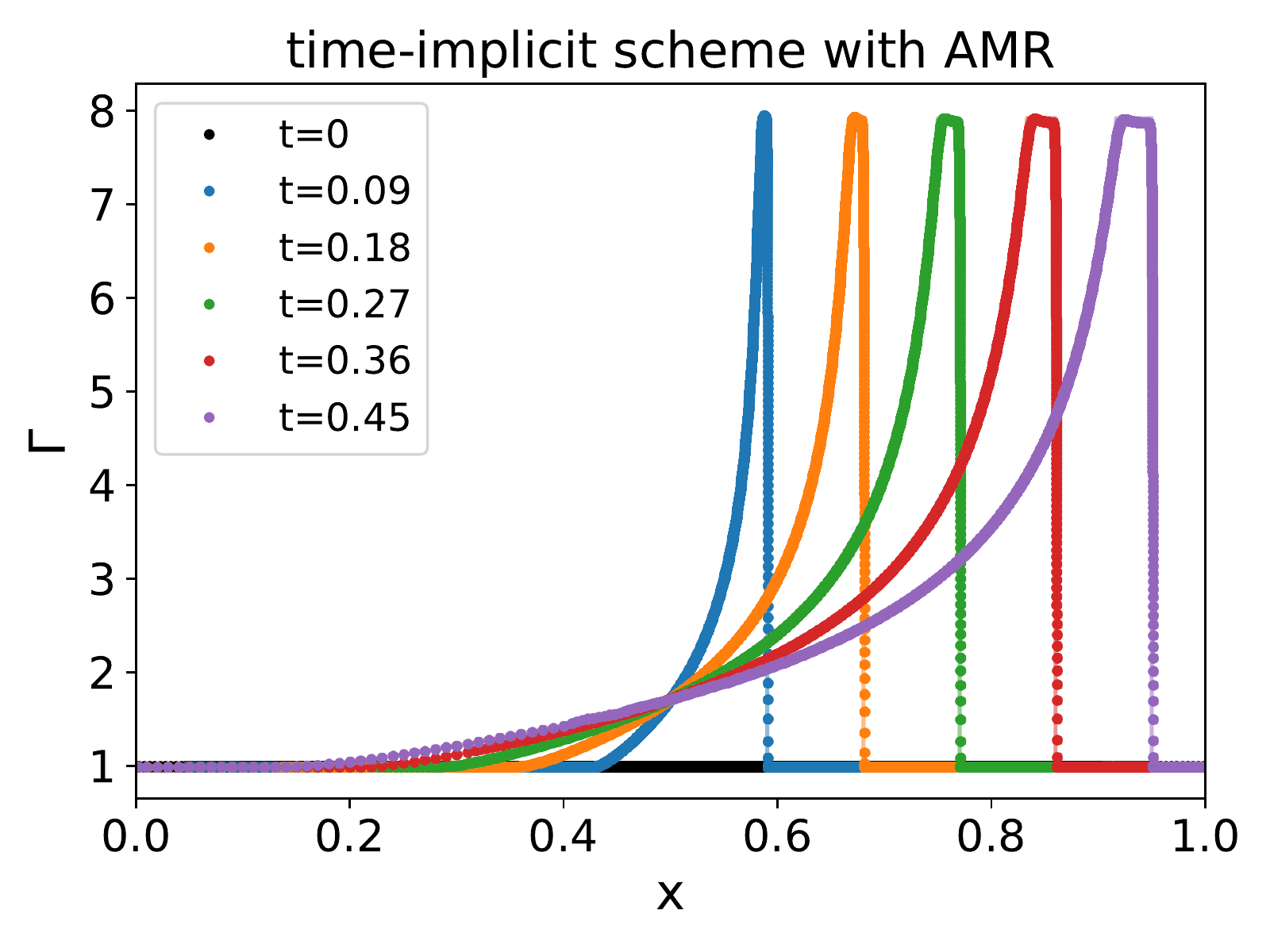}
\caption{
The spatial distributions of the relativistic density $D$ (upper panel) and of the Lorentz factor $\Gamma$ (lower panel) using the time-implicit solution procedure are shown. An initial pressure-jump $P_\mathrm{left}/P_\mathrm{right}=10^{5}$ and an initial density-jump of $\rho_\mathrm{left}/\rho_\mathrm{right}=10^{2}$ were used. A resolution of $N_\mathrm{AMR} \sim 20400$ was chosen and $s=0.5$. The other parameters were chosen as follows: $\alpha_\mathrm{art}=0.0$, $D_{\mathrm{diff,}D}=0.0$, $D_{\mathrm{diff,}M_x}=0.22$, $D_{\mathrm{diff,}{\cal E}^d} = 0.22$ and $\zeta = 0.5$. Note that in this calculation $\Gamma \sim 7.9$ has been reached, which reflects the extreme robustness and relatively good convergence of the method in the regime of very high Lorentz factors.
 \label{fig:HLF}}
\end{figure}
\begin{figure}
\includegraphics[width=\columnwidth]{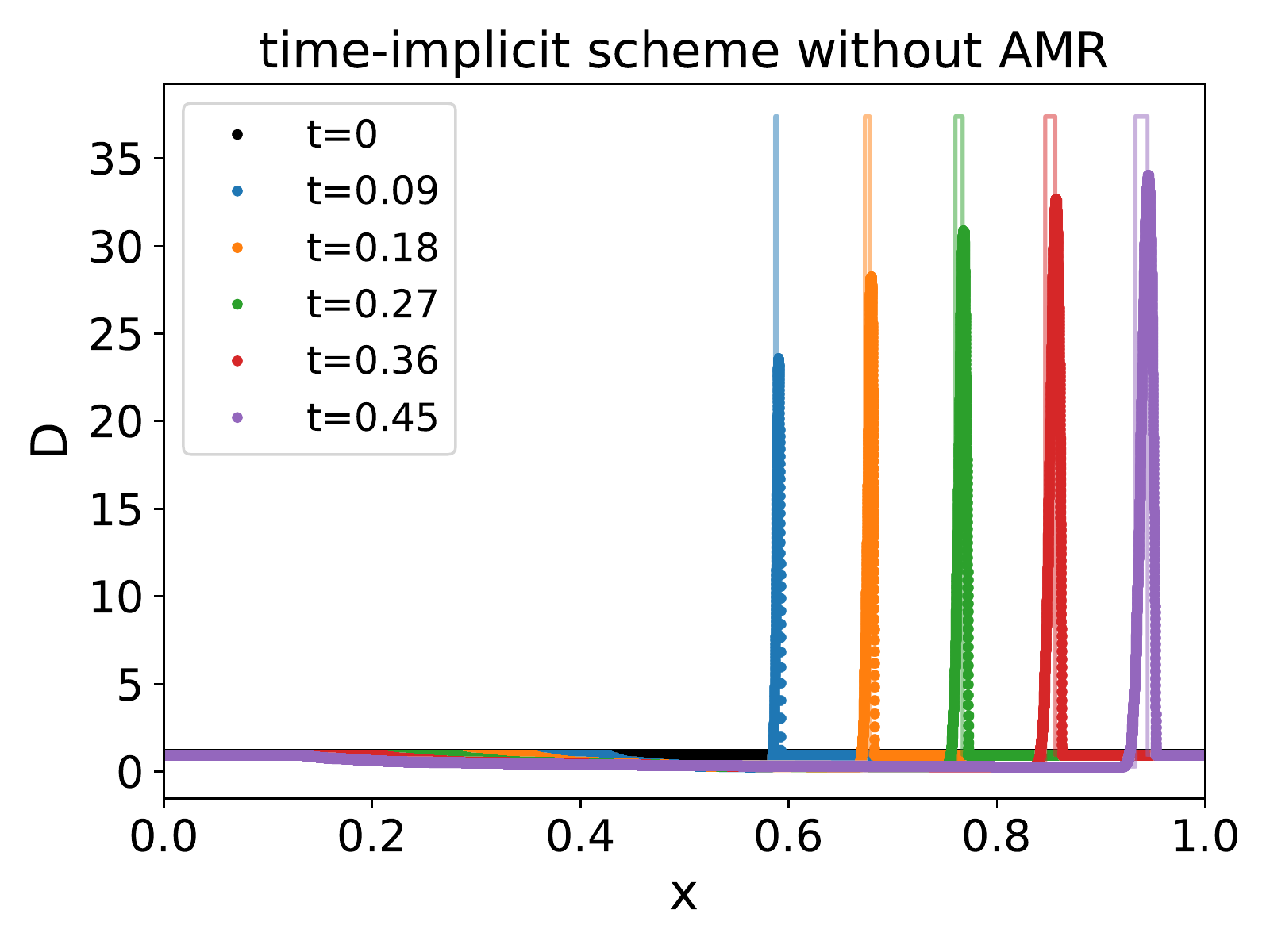}
\includegraphics[width=\columnwidth]{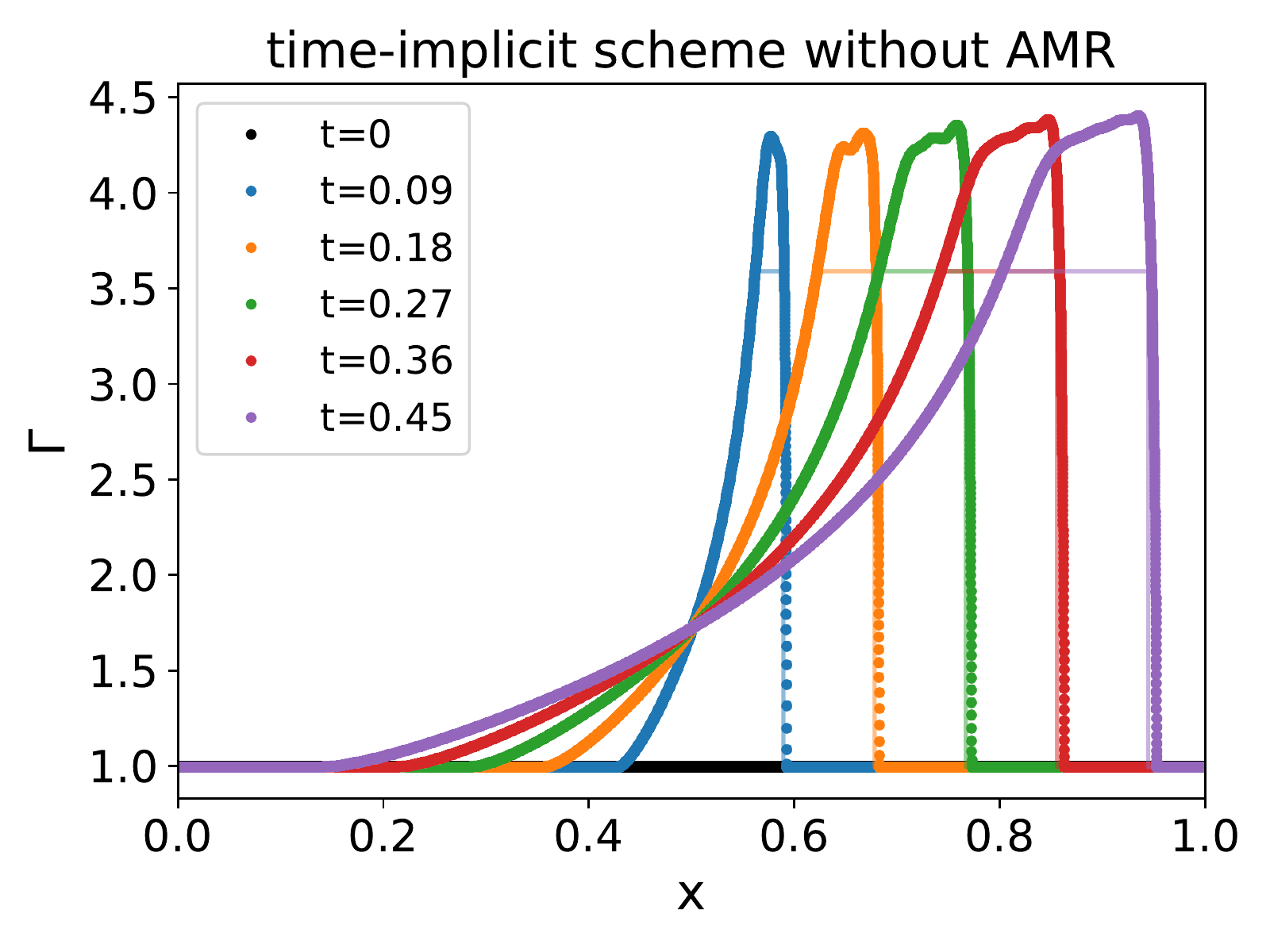}
\caption{
The spatial distributions of the relativistic density $D$ (upper panel) and the Lorentz factor $\Gamma$ (lower panel) using the time-implicit solution procedure without AMR are shown. An initial pressure and density jump of $P_\mathrm{left}/P_\mathrm{right}=10^{5}$, $\rho_\mathrm{left}/\rho_\mathrm{right}=1$ and $N_\mathrm{no~AMR} = 25600$ grid cells are used. The other parameters were chosen as follows: $\alpha_\mathrm{art}=0.0$, $D_{\mathrm{diff,}D}=0.25$, $D_{\mathrm{diff,}M_x}=0.25$, $D_{\mathrm{diff,}{\cal E}^d} = 0.25$ and $\zeta = 1.0$. The time step size has been increased during the simulation. Initially, $s = 0.5$ was chosen and increased after each output by $0.8$. Respectively the simulation has been run with $s=3.7$ between $t=0.36$ and $t=0.45$. Using the Lorentz factor of the exact solution this corresponds to a Courant number of ${\cal C}_\mathrm{CFL} \sim 3.6$. The ability to relax the Courant--Friedrichs--Lewy condition demonstrates the implicitness of our scheme. \label{fig:HCFL}}
\end{figure}

\begin{figure}
\includegraphics[width=\columnwidth]{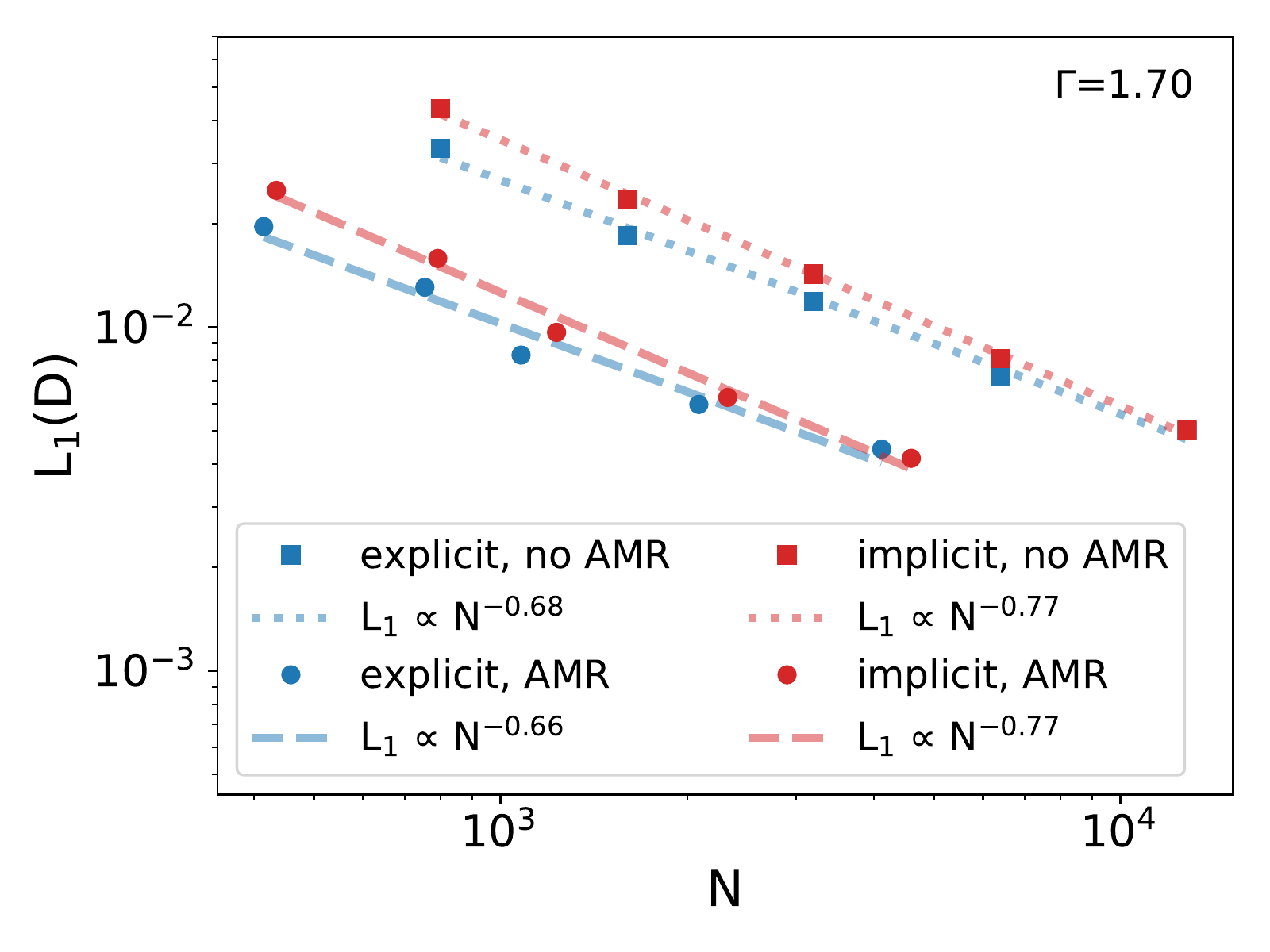}
\includegraphics[width=\columnwidth]{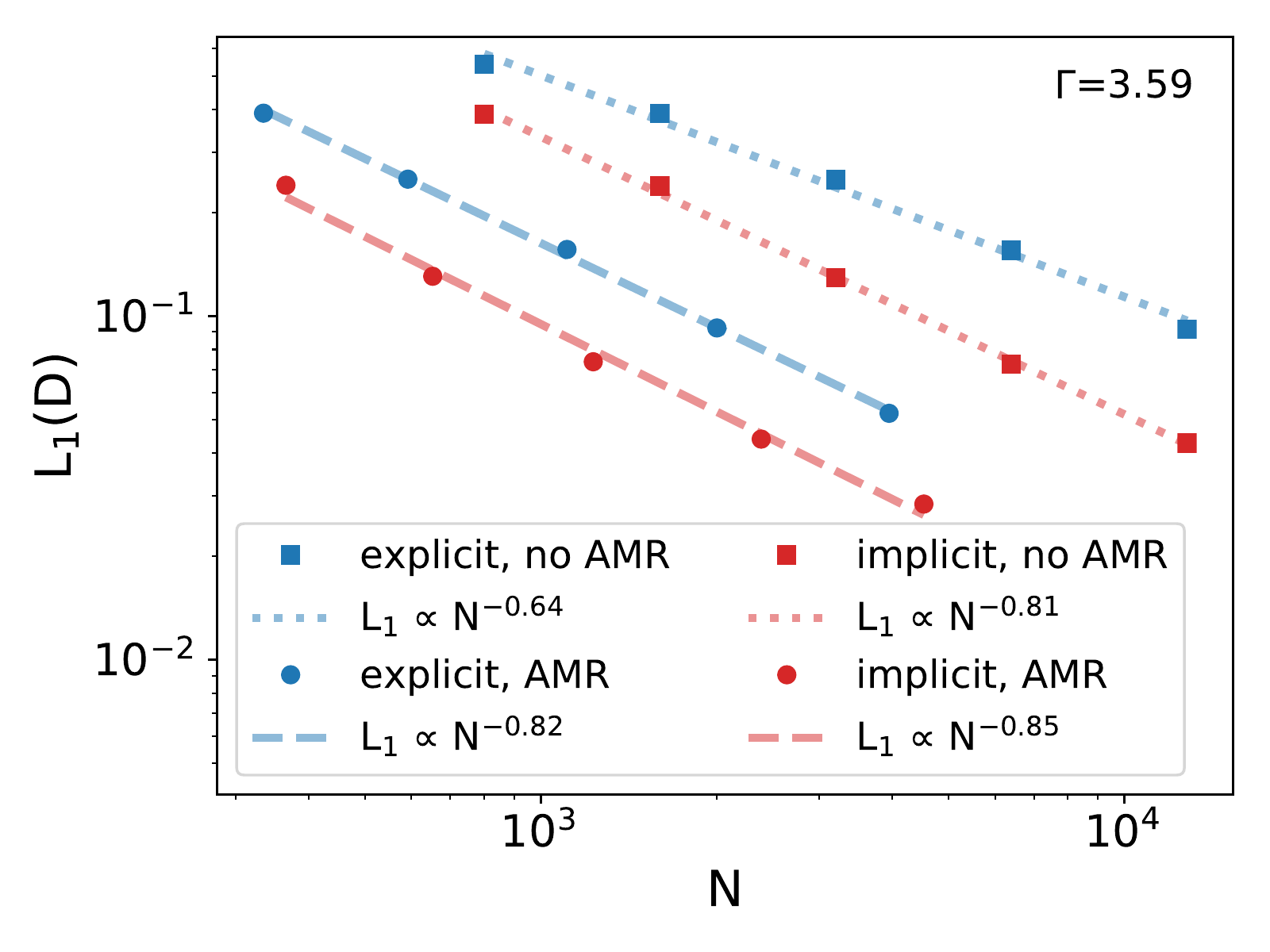}
\includegraphics[width=\columnwidth]{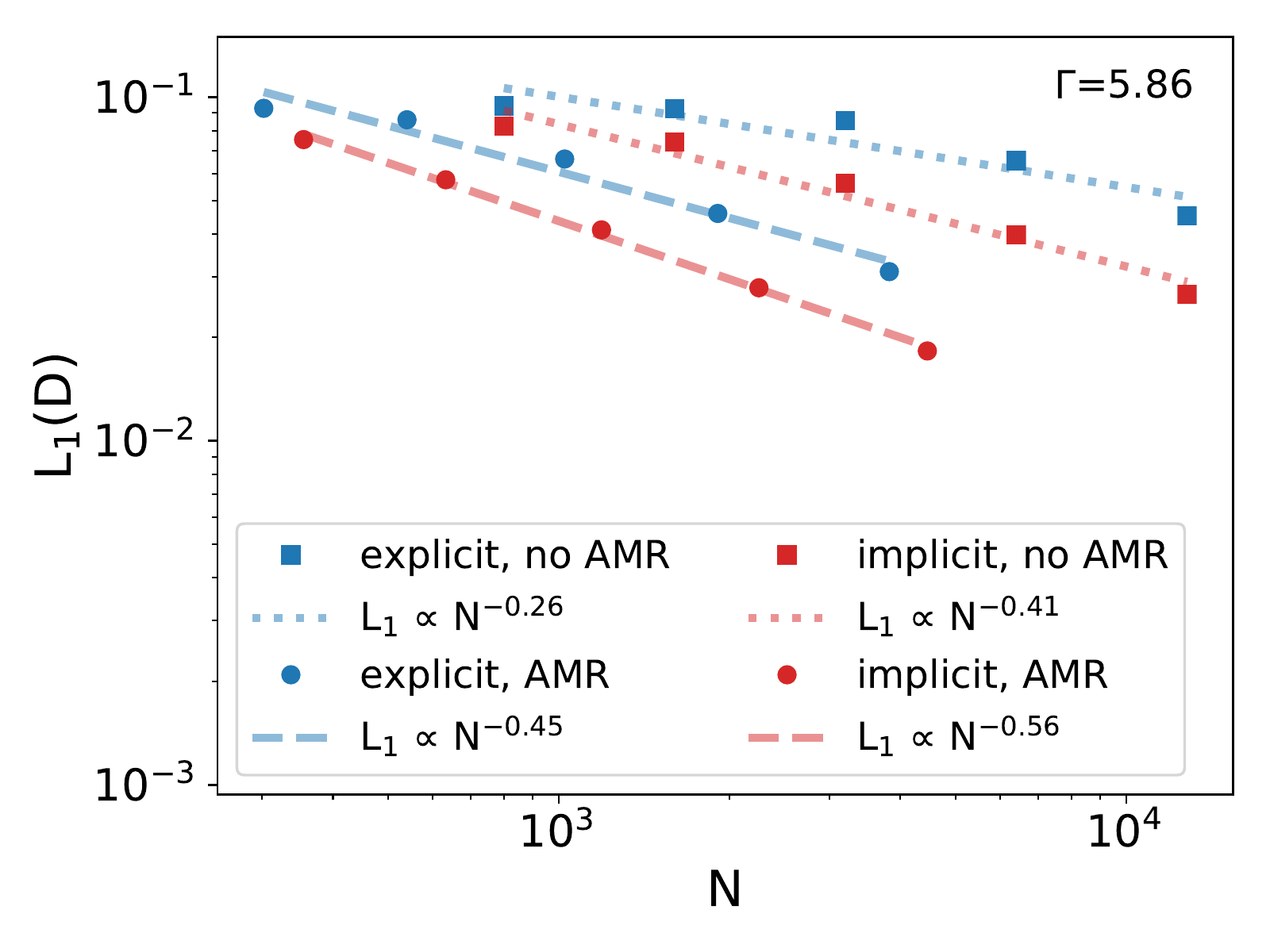}

\caption{$\mathsf{L_1}-$error norm computed from the relativistic density $D$ plotted as a function of the number of grid points. The results from the explicit/implicit scheme are displayed with and without AMR. The upper panel shows results from runs with a maximum Lorentz factor of $\Gamma = 1.70$, initial conditions are: $P_\mathrm{left}/P_\mathrm{right}= 10^2$, $\rho_\mathrm{left}/\rho_\mathrm{right}=1$. The middle panel shows results from runs with a maximum Lorentz factor of $\Gamma = 3.59$, initial conditions are: $P_\mathrm{left}/P_\mathrm{right}=10^5$, $\rho_\mathrm{left}/\rho_\mathrm{right}=1$. The lower panel shows results from runs with a maximum Lorentz factor of $\Gamma = 5.86$, initial conditions are: $P_\mathrm{left}/P_\mathrm{right}=10^5$, $\rho_\mathrm{left}/\rho_\mathrm{right}=10$. Moreover, the convergence rate was determined by fitting the plotted data. The chosen correction terms are described in section~\ref{sec:con_eff}. \label{fig:convergence}}
\end{figure}

\begin{figure}
\includegraphics[width=\columnwidth]{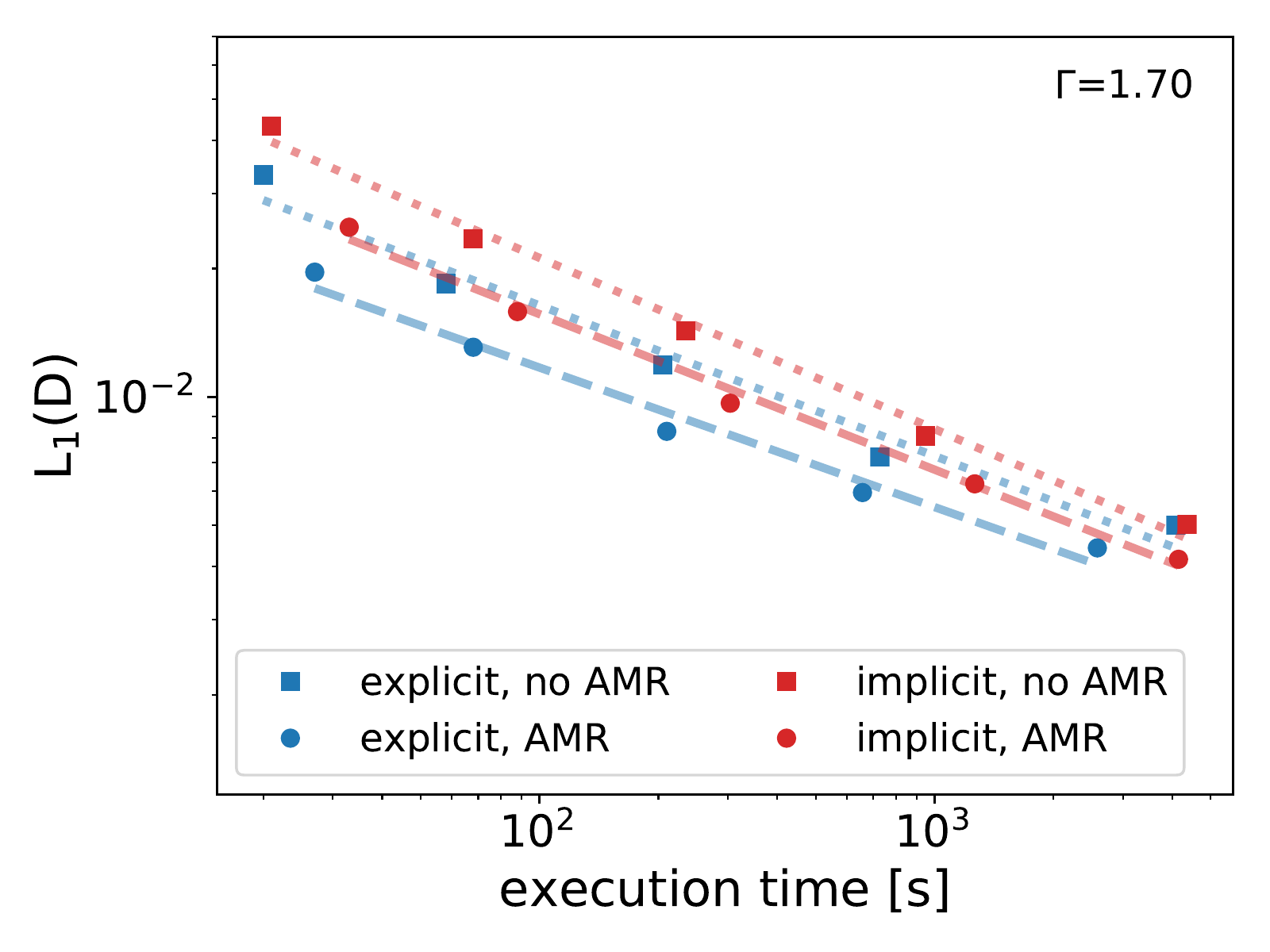}
\includegraphics[width=\columnwidth]{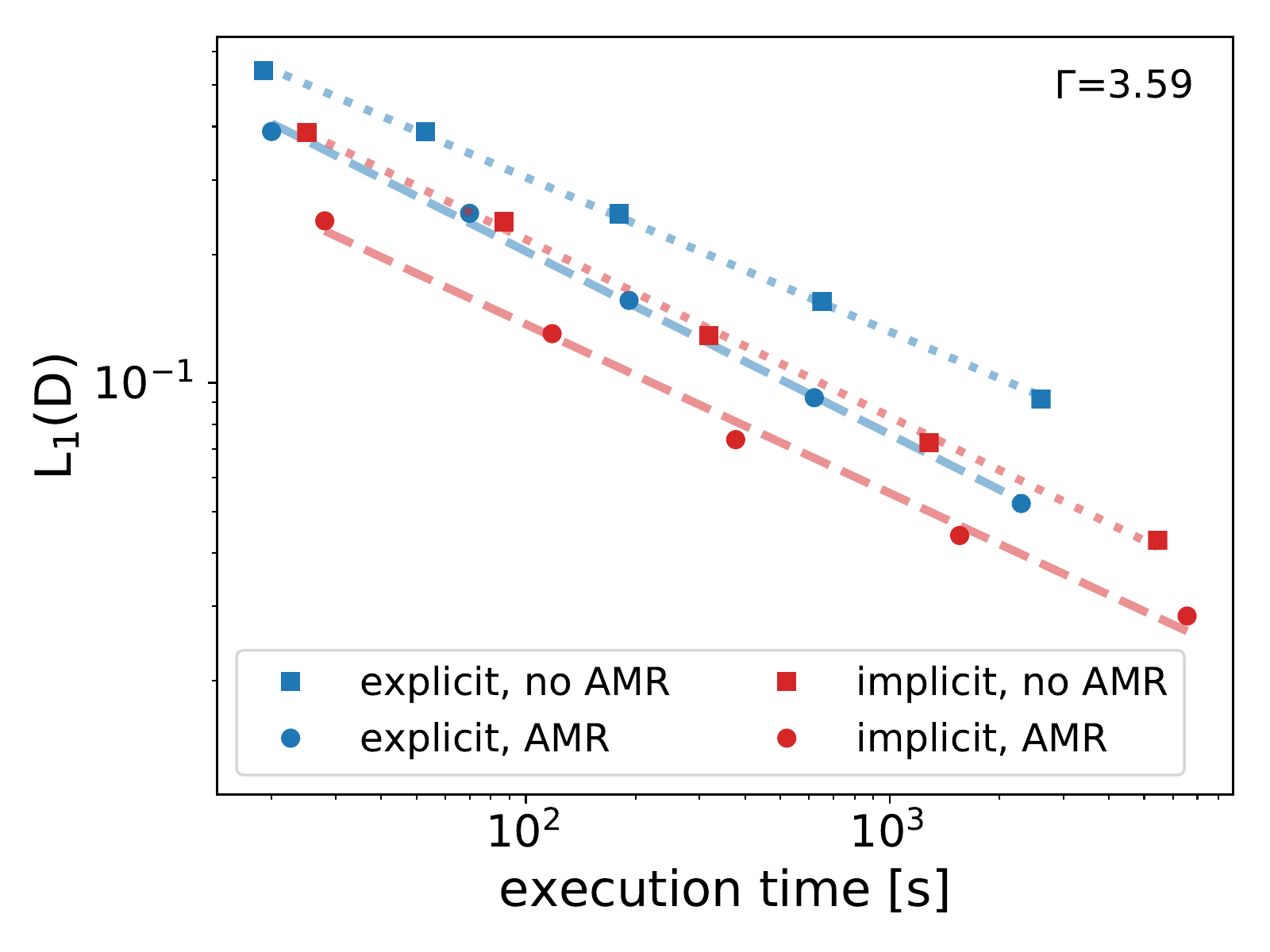}
\includegraphics[width=\columnwidth]{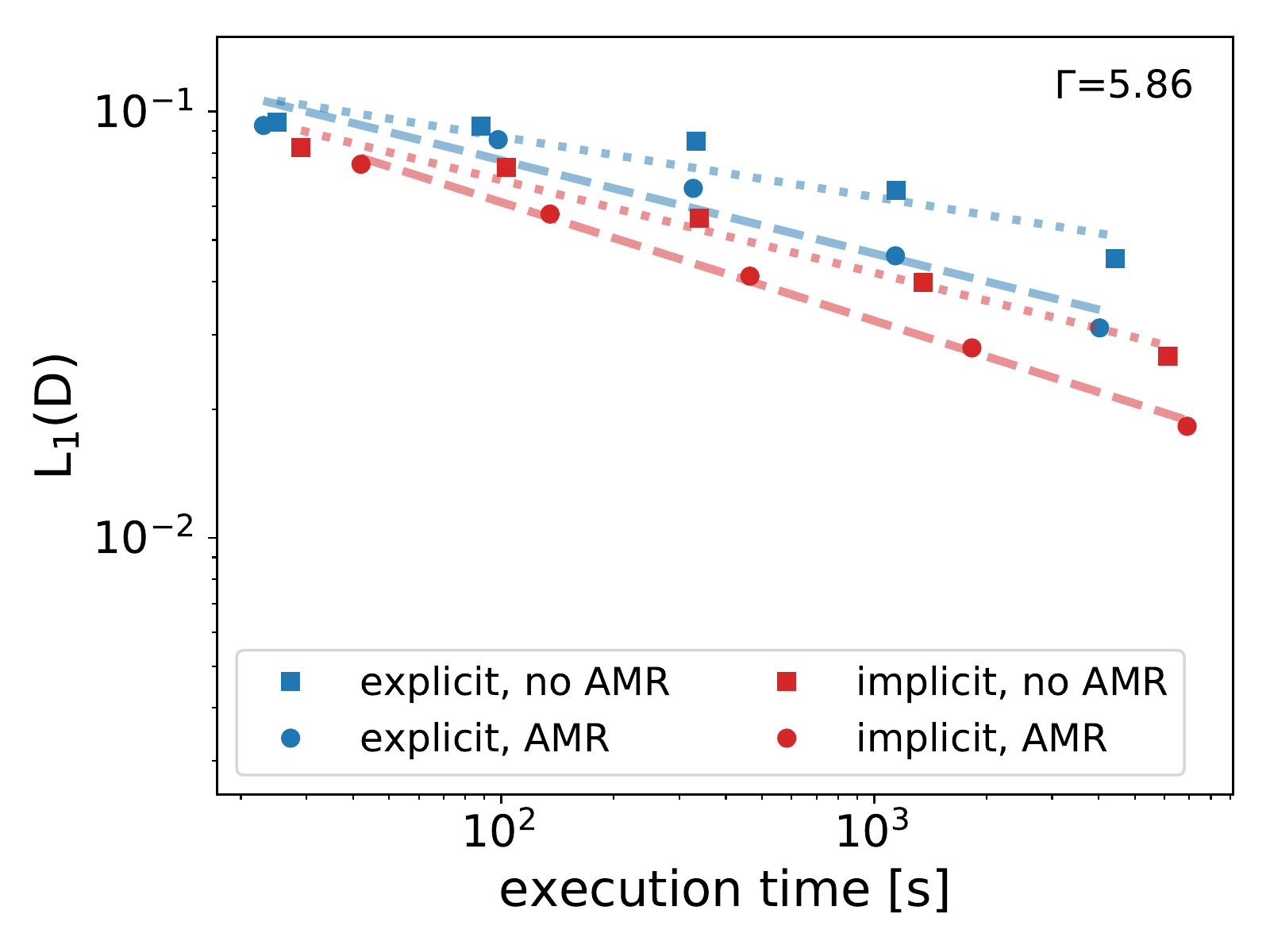}

\caption{$\mathsf{L_1-}$error norm computed from the relativistic density $D$ plotted as a function of the execution time. For the same runs as in Fig.~\ref{fig:convergence}.
\label{fig:time-accuracy}
}
\end{figure}

\begin{figure}
\centering
\includegraphics[width=\columnwidth]{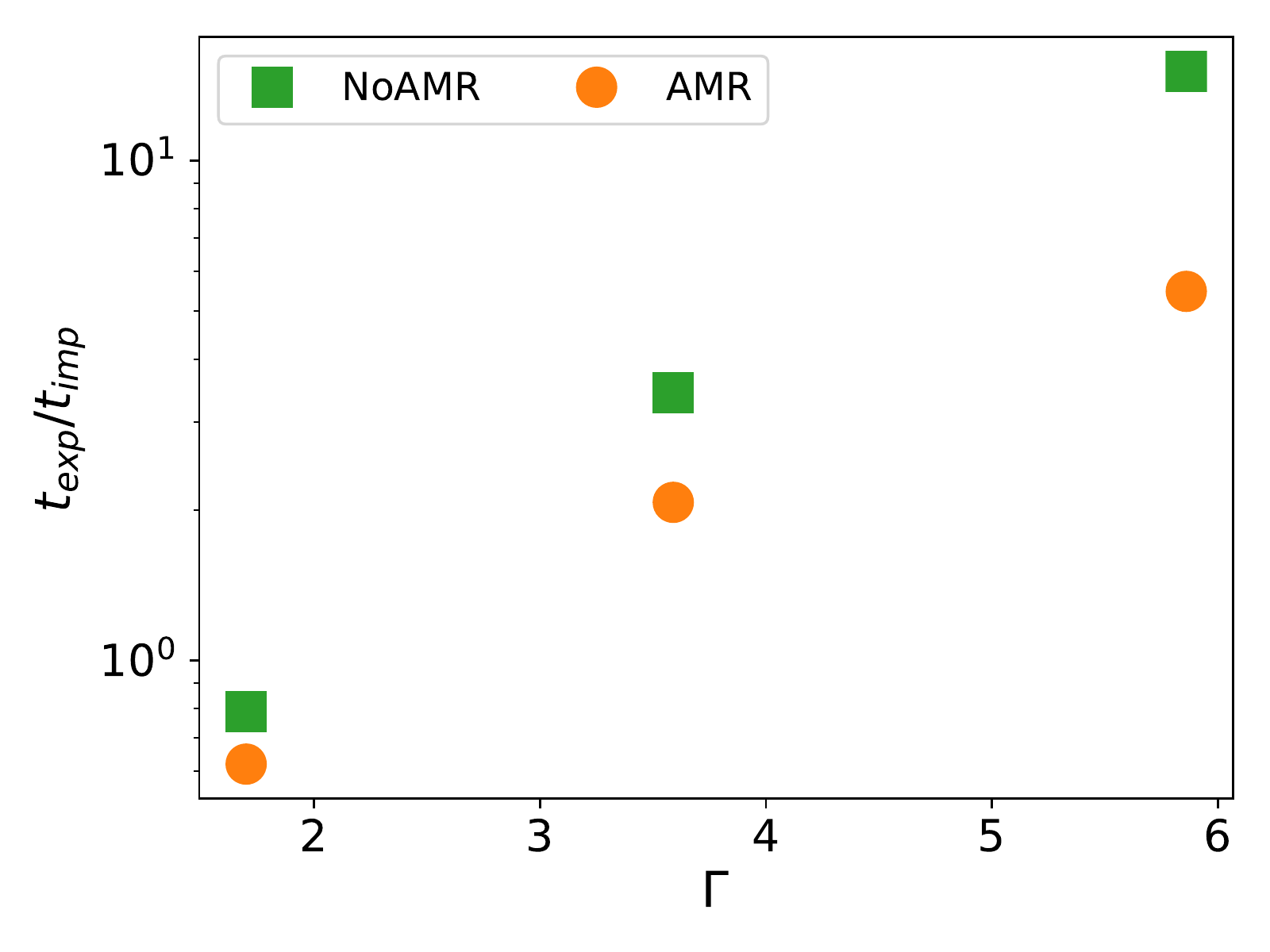}
\caption{The Quotient of the execution time of the implicit and explicit scheme as a function of Lorentz factor. The execution times for both schemes are estimated according to the fitted function in Figure \ref{fig:time-accuracy}. The execution times are the ones both schemes would require to reach the accuracy of the best resolved time-explicit simulations of Fig.~\ref{fig:time-accuracy}. This is evaluated for both, with and without AMR. We found that using AMR, the time-implicit scheme is $\sim 5.5$ times faster than the time-explicit scheme for the $\Gamma = 5.86$ case. Furthermore, the plot clearly shows that with increasing Lorentz factor the time-implicit scheme is becoming more and more superior. However, it is not superior in the regime of low Lorentz factors.
\label{fig:imp_vs_exp}}
\end{figure}


Furthermore, we used the explicit/implicit scheme with and without adaptive mesh refinement. The plots with $\Gamma = 1.70$ are shown in figures~\ref{fig:LF_r1_exp} and \ref{fig:LF_r1_imp}.
For both, the implicit and explicit scheme, we find that the numerical solution is closer to the exact one if adaptive mesh refinement is applied, although, the number of grid points is lower. The plots with $\Gamma = 3.59$ are shown in figures~\ref{fig:LF_r2_exp} and \ref{fig:LF_r2_imp}. For all these simulations correction terms were applied to reproduce the exact solution. Further information can be found in the captions of the corresponding figures. A comparison of the runs for the different Lorentz factors shows that problems with a higher Lorentz factor are numerically more difficult to solve accurately. As for low Lorentz factors, also at large values for $\Gamma$ the adaptive mesh refinement leads to a more accurate solution, although fewer grid points are used.

For the run of the lower panel of Fig.~\ref{fig:LF_r2_imp} (the time-implicit one with AMR for $\Gamma = 3.59$) we also show $D$ and ${\cal E}^d$ as a function of spatial position in Fig.~\ref{fig:p1_i2_AMR} and $P$ and $T$ are shown in Fig.~\ref{fig:p2_i2_AMR}. Especially the upper panel of Fig.~\ref{fig:p1_i2_AMR} demonstrates how powerful our implicit scheme is. Although only $N \sim 700$ grid points were used, the numerical solution of the relativistic density matches the exact one quite well at $t=0.45$. However, this depends on the width of the density pile up. It becomes broader when time is passing on. In fact, the major difficulty of these simulations is to reproduce the shock front with its narrow density peak. This becomes more difficult at higher Lorentz factors, as the peak becomes narrower.

\subsection{Very high Lorentz factors} \label{sec:very_high_LF}
Using the implicit scheme we are able to treat high Lorentz factors of at least $\Gamma \sim 7.9$ as shown in Fig.~\ref{fig:HLF}. Despite the Lorentz factor, which roughly matches the exact solution at $t = 0.45$, one can also notice in the upper panel that it is quite difficult to reproduce the relativistic density at the shock front. This is because the shock fronts are very narrow at high relativistic speeds and thus require a very high resolution. A similar run with the time-explicit scheme would be hardly possible. This is because with higher Lorentz factors the time-explicit scheme requires smaller values for $s = \Delta t/\Delta x$ or larger correction terms. Correspondingly, time-explicit runs with a high Lorentz factor become computationally very costly or are inaccurate, i.e. lacking from numerical artefacts.

\subsection{Computation with $C_\mathrm{CFL} > 1$}
Time-implicit schemes are known to be able to treat physical problems at higher CFL-numbers than time-explicit schemes. They are even able to relax the Courant--Friedrichs--Lewy condition.
In this section, we demonstrate that our time-implicit scheme can handle CFL-numbers much larger than one when adding an artificial diffusion term to the continuity, momentum and internal energy equation. Respective results are shown in Fig.~\ref{fig:HCFL}. We ran the corresponding simulation with $s = \Delta t / \Delta x$ chosen as an increasing function of time, such that the maximum is $s = 3.7$, which corresponds to a CFL-number of $\sim 3.6$. The discrepancy between the numerical and the exact solution for the relativistic density $D$ at the shock front is enhanced by the nonzero diffusion terms used. In contrast to our other simulations, we use it also in the continuity equation. This helps to maintain stability in this regime of high CFL-numbers. Furthermore, the deviation of the Lorentz factor could be reduced by adding artificial viscosity. Note, that for this run (in contrast to the other ones shown in this paper) the residual $R_j$ (see Eq.~\eqref{eq:residual}) was calculated by taking the diffusion terms into account.

\subsection{Convergence and efficiency} \label{sec:con_eff}
In this section, we outline several numerical properties of the solver as well as evaluate and compare the different numerical methods we have used. 
To measure the accuracy of the different schemes, we compute the $\mathsf{L_1}$ error norm:
\begin{equation}
\mathsf{L_1} = \frac{\sum^N_{j=1} \left|q_j-q(x_j)\right| \Delta x_j}{\sum^N_{j=1} \Delta x_j} \, ,
\end{equation}
where $q(x_j)$ denotes the exact solution at the same spatial position as of the numerical solution $q_j$.
This quantitative comparison is done in two steps, first, we study the accuracy as a function of the number of grid points as well as the convergence rate. Second, we investigate the accuracy as a function of execution time. This allows us to draw conclusions regarding the efficiency of both schemes. For these two steps, we compute the $\mathsf{L_1}$ error norm using the relativistic density distribution. Furthermore, we compare the results of three different initial conditions, which lead to disparate maximum Lorentz factors, i.e. $\Gamma = 1.70$, $\Gamma = 3.59$ and $\Gamma = 5.86$.
The first two test problems are the one presented in section~\ref{sec:spec_LF}. The third test problem we study leads to an even higher Lorentz factor of $\Gamma = 5.86$. The initial conditions are: $P_\mathrm{left}/P_\mathrm{right}=10^{5}$ and $\rho_\mathrm{left}/\rho_\mathrm{right}=10$. It is worth noting that the results depend on the chosen value for $s$. We have chosen the one that does roughly best in terms of producing the smallest $\mathsf{L_1}$ error for a given execution time. For the implicit scheme, we have chosen $s=0.5$, as previously mentioned. To maintain stability, the explicit scheme requires a much smaller value for $s$. According to this our choice is $s=0.01$ for all runs, independent of Lorentz factor.

The correction terms we have used for the simulations of this section are described below.
The runs with the lowest Lorentz factor ($\Gamma = 1.70$) were executed without an artificial viscosity term. However, diffusion was chosen for the implicit runs as: $D_{\mathrm{diff},D} = 0.0$, $D_{\mathrm{diff},M_x} = 0.1$ and $D_{\mathrm{diff},{\cal E}^d} = 0.1$. In the explicit runs the diffusion term was also involved but reduced with increasing resolution. For the lowest resolution runs the parameters are given by: $D_{\mathrm{diff},D} = 0.0$, $D_{\mathrm{diff},M_x} = 3.125 \cdot 10^{-6}$ and $D_{\mathrm{diff},{\cal E}^d} = 3.125 \cdot 10^{-6}$. Their values were halved each time the resolution was doubled. In addition to that, $\zeta = 0.2$ was used for the explicit runs.
Next, we describe the choice of parameters for the test problem which leads to the medium Lorentz factor ($\Gamma = 3.59$). The non-zero parameters for the implicit scheme are: $D_{\mathrm{diff},M_x} = 0.07$, $D_{\mathrm{diff},{\cal E}^d} = 0.07$ and $\zeta = 0.5$. For the explicit scheme the non-zero parameters are: $D_{\mathrm{diff},M_x} = 0.0001$, $D_{\mathrm{diff},{\cal E}^d} = 0.0001$, $\alpha_\mathrm{art}=6.0$ and $\zeta = 0.5$. Here again, the diffusion is given for the lowest resolution run and reduced for higher resolution as described above.
Last, we give the parameters of the runs, which involve the high Lorentz factor ($\Gamma = 5.86$). The non-zero parameters for the implicit scheme are: $D_{\mathrm{diff},M_x} = 0.17$, $D_{\mathrm{diff},{\cal E}^d} = 0.17$ and $\zeta = 0.57$. For the explicit scheme the parameters were chosen analogously to the previously mentioned ones and we give here the non-zero ones for the lowest resolution run: $D_{\mathrm{diff},M_x} = 0.00015$, $D_{\mathrm{diff},{\cal E}^d} = 0.00015$, $\alpha_\mathrm{art}=30.0$ and $\zeta = 1.0$.

Last it should be mentioned that for all runs the same number of threads, i.e. three threads, was used for the computation.

The results are displayed in figures~\ref{fig:convergence}--\ref{fig:time-accuracy} and discussed in the following. For the runs performed with the highest resolution the properties are given in Tab.~\ref{tab:ex_time}.\\ \\

\begin{table}
\caption{Explicit versus implicit methods: A list of runs using time-explicit and time-implicit solution strategy with/without AMR. Here $\Gamma,~N(\mathrm{t_{final}}), ~ \mathcal{C}_\mathrm{CFL},~\mathrm{time}$ and $\mathsf{L_1}$ refer to the Lorentz factor, the total number of cells at the end of the simulation, the corresponding CFL-number, the execution time and the $\mathsf{L_1}$-norm over the cells in the whole domain. The first 6-rows correspond to the runs with AMR (indicated by *), whereas the last 6-rows correspond to the runs without AMR and using 12800 uniformly distributed cells.
}
\label{tab:ex_time}
\begin{center}
\begin{tabular}{c|c|c|l|r|c}
\hline
\hline
$\Gamma$ & scheme & $N(\mathrm{t_{final}})$ & $\mathcal{C}_\mathrm{CFL}$ & time
[s] & $\mathsf{L_1}(\mathrm{t_{final}})$ \\ \hline
1.70 & explicit* & 4117 & 0.0081 & 2582 & $4.4 \cdot 10^{-3}$ \\
1.70 & implicit* & 4596 & 0.40 & 4144 & $4.2 \cdot 10^{-3}$ \\
3.59 & explicit* & 3951 & 0.0096 & 2297 & $ 5.2 \cdot 10^{-2}$ \\
3.59 & implicit* & 4532 & 0.48 & 6559 & $ 2.8 \cdot 10^{-2}$ \\
5.86 & explicit* & 3823 & 0.0099 & 4024 & $3.1 \cdot 10^{-2}$ \\
5.86 & implicit* & 4461 & 0.49 & 6907 & $1.8 \cdot 10^{-2}$ \\
\hline
1.70 & explicit & 12800 & 0.0081 & 4080 & $5.0 \cdot 10^{-3}$ \\
1.70 & implicit & 12800 & 0.40 & 4356 & $5.0 \cdot 10^{-3}$ \\
3.59 & explicit & 12800 & 0.0096 & 2602 & $ 9.2 \cdot 10^{-2}$ \\
3.59 & implicit & 12800 & 0.48 & 5447 & $ 4.3 \cdot 10^{-2} $ \\
5.86 & explicit & 12800 & 0.0099 & 4437 & $4.5 \cdot 10^{-2}$ \\
5.86 & implicit & 12800 & 0.49 & 6134 & $2.7 \cdot 10^{-2}$ \\
\hline
\hline
\end{tabular}
\end{center}
\end{table}
 
\subsubsection*{\textbf{Error estimates versus number of cells}}
We investigated the numerical error $\mathsf{L_1}$ as a function of the number of grid points used and determined the convergence rate. The results are shown in Fig.~\ref{fig:convergence}. We found that AMR clearly enhances the convergence rate of both schemes, time-implicit and explicit. Interestingly, the numerical error of the time-implicit scheme and the time-explicit scheme are close together. This can be an incident of the chosen value for $s = \Delta t / \Delta x$.
Although a priori error estimates of a discretized set of equations are generally lower than the corresponding a posteriori ones, we expect the convergence rates in model problems dominated by the propagation of relativistic shock fronts to be even much lower due to the strong non-linearities and nearly singular behaviour of the advection terms across the shock fronts. However, the convergence rate for the runs with the medium Lorentz factor appears to be higher than for the problem leading to the lowest Lorentz factor, though the averaged point-wise errors are comparatively larger when using a fixed number of grid points. The convergence rates for the time-explicit and implicit schemes seem to be similar when using AMR. Nevertheless, the error for the implicit scheme is lower for this test problem. For the highest Lorentz factor studied here, the convergence rate is lowest. We want to point out that in this regime, the implicit scheme is more accurate and converges faster than the explicit one.

\subsubsection*{\textbf{Error estimates versus execution time}}
Here we investigate the numerical error $\mathsf{L_1}$ as function of the execution time. The results are shown in Fig.~\ref{fig:time-accuracy}. We found that the use of AMR speeds up the computation for a desired accuracy. For the runs with the lower Lorentz factor, the time-explicit and implicit schemes are most similar, compared to the runs involving higher Lorentz factors. Here, the time-explicit scheme is more efficient. It requires less computation time to achieve a given accuracy. However, this is not true for high Lorentz factors, as the lower panels demonstrate. The accuracy of the time-implicit scheme is significantly better than the one of the explicit scheme. Especially, we want to point out that the rate, by which the accuracy increases with computation time is much better for the implicit scheme than for the explicit one in the regime of ultra-relativistic shock fronts. This is demonstrated by the lower panel.

We evaluated the differences in efficiency by computing the speedup of the implicit scheme over the explicit one. The results are shown in Fig.~\ref{fig:imp_vs_exp}. The caption gives the computational details. The figure clearly demonstrates that the time-implicit scheme becomes much more efficient with and without AMR in the limit of ultra-relativistic shock fronts.

However, one should be aware that the absolute values of the $\mathsf{L_1}$ error norm can't be directly compared between the two schemes to derive a general statement about explicit and implicit relativistic hydrodynamics. Nevertheless, the found trend states that the implicit approach becomes better compared to the explicit one with increasing Lorentz factor. We are confident that this trend can also be found when other schemes for relativistic hydrodynamics are compared.
\section{Summary and Conclusions} \label{sec:conclusions}
In this paper, we have presented a unified numerical approach for modelling the propagation of ultra-relativistic shocks within the framework of the pre-conditioned defect-correction iteration procedure. Our numerical solver relies on the finite volume formulation to enhance physical consistency and ensures the conservation of mass, momentum and energy. The momentum is computed using a staggered grid. For achieving high spatial accuracy we have adopted a formulation using flux limiters, whereas the discretization used in the time-implicit scheme is second-order temporal accurate. The defect-correction iteration procedure is employed with preconditionings that correspond to first-order spatial accuracy or the identity matrix. While in the former case the numerical procedure is capable of CFL-numbers larger than unity, the later one can only treat flows at $C_\mathrm{CFL}<1$. This is commonly referred to as unconditionally and conditional stable. In practice, the stability for each of the two schemes can only be obtained for a well-chosen set of values for the parameters of the correction terms.

For boosting efficiency, the numerical code has been parallelised using domain decomposition and made capable of adaptive mesh refinement for dynamically increasing the grid density in the regions of interest.

The numerical tests performed here have shown differences between the time-implicit and time-explicit solution procedure. Our main results are:
\begin{itemize}
\item In the regime of high Lorentz factors, i.e. $\Gamma \gtrsim 3$, the time-implicit numerical solver is found to be much more accurate and efficient than its time-explicit counterpart. This difference between the time-explicit and time-implicit scheme becomes even stronger with an increasing Lorentz factor. 

\item The time-implicit solver is capable of modelling the propagation of high-relativistic shock fronts with $\mathcal{C}_\mathrm{CFL}$ larger than unity.

\item Generally speaking, time-implicit solvers may become computationally superior over time-explicit ones in the regime of very high Lorentz factors.
Here the very strong non-linearity is challenging for an explicit formulation. We found an implicit procedure to be more capable of this.

\end{itemize}
We expect this superiority to be more obvious if the concerned plasma is non-ideal, dissipative, magnetized, radiative and includes chemical processes that operate at much shorter time scales than the dynamical one.

Finally, we note that our code can be used interactively via a webpage\footnote[1]{\url{https://typo.iwr.uni-heidelberg.de/groups/compastro/computer-codes-and-numerical-solvers/rstp/} (see C++ based solver)}.

\section*{Acknowledgements}
We are thankful for carrying out the simulations at the compute server of the Interdisciplinary Center for Scientific Computing (IWR) of Heidelberg University.



\bibliographystyle{mnras}
\bibliography{references} 

\begin{thebibliography}{}
\makeatletter
\relax
\def\mn@urlcharsother{\let\do\@makeother \do\$\do\&\do\#\do\^\do\_\do\%\do\~}
\def\mn@doi{\begingroup\mn@urlcharsother \@ifnextchar [ {\mn@doi@}
  {\mn@doi@[]}}
\def\mn@doi@[#1]#2{\def\@tempa{#1}\ifx\@tempa\@empty \href
  {http://dx.doi.org/#2} {doi:#2}\else \href {http://dx.doi.org/#2} {#1}\fi
  \endgroup}
\def\mn@eprint#1#2{\mn@eprint@#1:#2::\@nil}
\def\mn@eprint@arXiv#1{\href {http://arxiv.org/abs/#1} {{\tt arXiv:#1}}}
\def\mn@eprint@dblp#1{\href {http://dblp.uni-trier.de/rec/bibtex/#1.xml}
  {dblp:#1}}
\def\mn@eprint@#1:#2:#3:#4\@nil{\def\@tempa {#1}\def\@tempb {#2}\def\@tempc
  {#3}\ifx \@tempc \@empty \let \@tempc \@tempb \let \@tempb \@tempa \fi \ifx
  \@tempb \@empty \def\@tempb {arXiv}\fi \@ifundefined
  {mn@eprint@\@tempb}{\@tempb:\@tempc}{\expandafter \expandafter \csname
  mn@eprint@\@tempb\endcsname \expandafter{\@tempc}}}

\bibitem[\protect\citeauthoryear{{Brezinski} \& {Hujeirat}}{{Brezinski} \&
  {Hujeirat}}{2011}]{Brezinski_2011}
{Brezinski} F.,  {Hujeirat} A.~A.,  2011, \mn@doi [Astronomy Studies
  Development] {10.4081/ads.2011.e4}, \href
  {https://ui.adsabs.harvard.edu/\#abs/2011AstSD...1....4B} {1, 4}

\bibitem[\protect\citeauthoryear{{Commer{\c c}on}, {Debout}  \&
  {Teyssier}}{{Commer{\c c}on} et~al.}{2014}]{Commercon_2014}
{Commer{\c c}on} B.,  {Debout} V.,   {Teyssier} R.,  2014, \mn@doi [\aap]
  {10.1051/0004-6361/201322858}, \href
  {http://adsabs.harvard.edu/abs/2014A\%26A...563A..11C} {563, A11}

\bibitem[\protect\citeauthoryear{{Fromang}, {Hennebelle}  \&
  {Teyssier}}{{Fromang} et~al.}{2006}]{Fromang_2006}
{Fromang} S.,  {Hennebelle} P.,   {Teyssier} R.,  2006, \mn@doi [\aap]
  {10.1051/0004-6361:20065371}, \href
  {http://adsabs.harvard.edu/abs/2006A\%26A...457..371F} {457, 371}

\bibitem[\protect\citeauthoryear{G{\'o}mez et~al.,}{G{\'o}mez
  et~al.}{2016}]{Gomez_2016}
G{\'o}mez J.~L.,  et~al., 2016, The Astrophysical Journal, 817, 96

\bibitem[\protect\citeauthoryear{Hackbusch}{Hackbusch}{1994}]{Hackbusch_1994}
Hackbusch W.,  1994, Iterative Solution of Large Sparse Systems of Equations.
Applied mathematical sciences, Springer-Verlag, \url
  {https://books.google.de/books?id=xu0ZAQAAIAAJ}

\bibitem[\protect\citeauthoryear{{Harten} \& {Osher}}{{Harten} \&
  {Osher}}{1987}]{Harten_1987}
{Harten} A.,  {Osher} S.,  1987, \mn@doi [SIAM Journal on Numerical Analysis]
  {10.1137/0724022}, \href
  {https://ui.adsabs.harvard.edu/abs/1987SJNA...24..279H} {24, 279}

\bibitem[\protect\citeauthoryear{{Hujeirat}}{{Hujeirat}}{2005a}]{Hujeirat_2005}
{Hujeirat} A.,  2005a, \mn@doi [\na] {10.1016/j.newast.2004.09.003}, \href
  {https://ui.adsabs.harvard.edu/\#abs/2005NewA...10..173H} {10, 173}

\bibitem[\protect\citeauthoryear{{Hujeirat}}{{Hujeirat}}{2005b}]{Hujeirat_2005CoPhC}
{Hujeirat} A.,  2005b, \mn@doi [Computer Physics Communications]
  {10.1016/j.cpc.2005.01.013}, \href
  {https://ui.adsabs.harvard.edu/\#abs/2005CoPhC.168....1H} {168, 1}

\bibitem[\protect\citeauthoryear{{Hujeirat} \& {Thielemann}}{{Hujeirat} \&
  {Thielemann}}{2009a}]{Hujeirat_2009Spectrum}
{Hujeirat} A.~A.,  {Thielemann} F.~K.,  2009a, Informatik-Spektrum, \href
  {https://ui.adsabs.harvard.edu/\#abs/2009InfSp..32..496H} {32, 496}

\bibitem[\protect\citeauthoryear{{Hujeirat} \& {Thielemann}}{{Hujeirat} \&
  {Thielemann}}{2009b}]{Hujeirat_2009LowMach}
{Hujeirat} A.~A.,  {Thielemann} F.~K.,  2009b, \mn@doi [\mnras]
  {10.1111/j.1365-2966.2009.15498.x}, \href
  {https://ui.adsabs.harvard.edu/\#abs/2009MNRAS.400..903H} {400, 903}

\bibitem[\protect\citeauthoryear{{Hujeirat}, {Livio}, {Camenzind}  \&
  {Burkert}}{{Hujeirat} et~al.}{2003}]{Hujeirat_2003HLCB}
{Hujeirat} A.,  {Livio} M.,  {Camenzind} M.,   {Burkert} A.,  2003, \mn@doi
  [\aap] {10.1051/0004-6361:20031040}, \href
  {https://ui.adsabs.harvard.edu/\#abs/2003A&A...408..415H} {408, 415}

\bibitem[\protect\citeauthoryear{Khokhlov}{Khokhlov}{1998}]{Khokhlov_1998}
Khokhlov A.,  1998, \mn@doi [Journal of Computational Physics]
  {https://doi.org/10.1006/jcph.1998.9998}, 143, 519

\bibitem[\protect\citeauthoryear{Kurganov \& Tadmor}{Kurganov \&
  Tadmor}{2002}]{Kurganov_2002}
Kurganov A.,  Tadmor E.,  2002, \mn@doi [Numerical Methods for Partial
  Differential Equations] {10.1002/num.10025}, 18, 584

\bibitem[\protect\citeauthoryear{{Lightman}, {Press}, {Price}  \&
  {Teukolsky}}{{Lightman} et~al.}{1975}]{Lightman_1975}
{Lightman} A.~P.,  {Press} W.~H.,  {Price} R.~H.,   {Teukolsky} S.~A.,  1975,
  {Problem Book in Relativity and Gravitation}.
Princeton University Press

\bibitem[\protect\citeauthoryear{Mart{\'i} \& M{\"u}ller}{Mart{\'i} \&
  M{\"u}ller}{2003}]{Marti_2003}
Mart{\'i} J.~M.,  M{\"u}ller E.,  2003, \mn@doi [Living Reviews in Relativity]
  {10.12942/lrr-2003-7}, 6, 7

\bibitem[\protect\citeauthoryear{Mart{\'i} \& M{\"u}ller}{Mart{\'i} \&
  M{\"u}ller}{2015}]{Marti_2015}
Mart{\'i} J.~M.,  M{\"u}ller E.,  2015, \mn@doi [Living Reviews in
  Computational Astrophysics] {10.1007/lrca-2015-3}, 1, 3

\bibitem[\protect\citeauthoryear{{Norman} \& {Winkler}}{{Norman} \&
  {Winkler}}{1986}]{Norman_1986}
{Norman} M.~L.,  {Winkler} K.-H.~A.,  1986, in {Winkler} K.-H.~A.,  {Norman}
  M.~L.,  eds,  NATO Advanced Science Institutes (ASI) Series C Vol. 188, NATO
  Advanced Science Institutes (ASI) Series C. p.~449

\bibitem[\protect\citeauthoryear{Sod}{Sod}{1978}]{Sod_1978}
Sod G.~A.,  1978, \mn@doi [Journal of Computational Physics]
  {https://doi.org/10.1016/0021-9991(78)90023-2}, 27, 1

\bibitem[\protect\citeauthoryear{Sweby}{Sweby}{1984}]{Sweby_1984}
Sweby P.,  1984, SIAM Journal of Numerical Analysis, 21, 995

\bibitem[\protect\citeauthoryear{{Teyssier}}{{Teyssier}}{2002}]{Teyssier_2002}
{Teyssier} R.,  2002, \mn@doi [\aap] {10.1051/0004-6361:20011817}, \href
  {http://adsabs.harvard.edu/abs/2002A\%26A...385..337T} {385, 337}

\bibitem[\protect\citeauthoryear{{Thompson}}{{Thompson}}{1985}]{Thompson_1985}
{Thompson} K.~W.,  1985, PhD thesis, Princeton Univ., NJ.

\bibitem[\protect\citeauthoryear{Thompson}{Thompson}{1986}]{Thompson_1986}
Thompson K.~W.,  1986, \mn@doi [Journal of Fluid Mechanics]
  {10.1017/S0022112086001489}, 171, 365–375

\bibitem[\protect\citeauthoryear{{Wilson}}{{Wilson}}{1972}]{Wilson_1972}
{Wilson} J.~R.,  1972, \mn@doi [\apj] {10.1086/151434}, \href
  {http://adsabs.harvard.edu/abs/1972ApJ...173..431W} {173, 431}

\bibitem[\protect\citeauthoryear{{van Leer}}{{van Leer}}{1979}]{vanLeer_1979}
{van Leer} B.,  1979, \mn@doi [Journal of Computational Physics]
  {10.1016/0021-9991(79)90145-1}, \href
  {https://ui.adsabs.harvard.edu/abs/1979JCoPh..32..101V} {32, 101}

\makeatother
\end{thebibliography}






\bsp	
\label{lastpage}
\end{document}